\documentclass[fleqn,usenatbib]{mnras}

\usepackage{fix-cm}

\usepackage{newtxtext,newtxmath}
\usepackage{diagbox}   
\usepackage{booktabs}  

\usepackage[T1]{fontenc}

\DeclareRobustCommand{\VAN}[3]{#2}
\let\VANthebibliography\thebibliography
\def\thebibliography{\DeclareRobustCommand{\VAN}[3]{##3}\VANthebibliography}

\usepackage{graphicx}
\usepackage{amsmath}
\usepackage{xcolor}
\usepackage[normalem]{ulem}

\newcommand{\cnnfof}{CNN+FoF}

\title[\cnnfof{}: deep learning for halo identification]{\cnnfof{}: application of deep learning to the identification of dark matter haloes}

\author[S. Maiti et al.]{
Soumadeep Maiti,$^{1,2}$\thanks{E-mail: maitisoumadeep@gmail.com}
Carlos M. Correa,$^{2}$
Andrea Fiorilli,$^{2}$
Andrés N. Ruiz,$^{3,4}$
Dante J. Paz,$^{3,4}$
\newauthor
Alejandro Pérez-Fernández,$^{2}$
Ariel G. Sánchez$^{1,2}$
\\
$^{1}$Ludwig-Maximilians-Universität München, Geschwister-Scholl-Platz 1
80539 München, Germany\\
$^{2}$Max Planck Institute for Extraterrestrial Physics, Giessenbachstr. 1, 85748 Garching, Germany\\
$^{3}$Instituto de Astronom\'{\i}a Te\'orica y Experimental, UNC--CONICET, Laprida 854, X5000BGR C\'ordoba, Argentina\\
$^{4}$Observatorio Astron\'omico de C\'ordoba, UNC, Laprida 854, X5000BGR C\'ordoba, Argentina
}

\date{Accepted XXX. Received YYY; in original form ZZZ}
\pubyear{\the\year{}}

\begin{document}
\label{firstpage}
\pagerange{\pageref{firstpage}--\pageref{lastpage}}
\maketitle

\begin{abstract}
We present a deep-learning-based approach for identifying dark matter haloes in cosmological N-body simulations.
Our framework consists of a volumetric Convolutional Neural Network to classify individual simulation particles as either halo or non-halo members, followed by a highly optimised and parallelised Friends-of-Friends clustering algorithm that groups the classified halo members into distinct haloes.
The training data comprise simulations generated using \texttt{GADGET-4}, with labels obtained with the \texttt{ROCKSTAR} halo finder. Our models incorporate two main halo mass definitions, $M_{200\mathrm{b}}$ and $M_{\text{vir}}$, with similar performance. 
For haloes defined by the \texttt{ROCKSTAR} $M_{200\mathrm{b}}$ criterion, the classification network demonstrated stable performance across multiple simulation resolutions.
For the highest resolution, it achieved over $98\%$ across all primary performance metrics when identifying halo particles. 
Furthermore, the FoF algorithm yielded halo catalogues with a purity generally exceeding $99\%$ and a stable completeness of $94\%$ for masses above $5\times10^{11} \, M_\odot$.
Our pipeline recovered the centre-of-mass positions, velocities and halo masses with high fidelity, yielding a halo mass function consistent to within $5\%$ of the reference while faithfully reconstructing the internal density profiles.
The primary objective of this study is to offer a faster and scalable alternative to conventional halo finders, achieving a speed-up of approximately one order of magnitude relative to \texttt{ROCKSTAR}. While traditional N-body simulations remain dominated by the time required to solve gravitational dynamics, this post-processing acceleration becomes essential when coupled with fast field-level emulators, offering a promising pathway towards efficient, end-to-end forward modelling in modern simulation-based inference.
\end{abstract}



\section{Introduction}
\label{sec:intro}

The Lambda cold dark matter ($\Lambda$CDM) model is the standard theoretical framework for describing the gravitational evolution of the large-scale structure of the Universe from small initial density perturbations. Numerical N-body simulations based on this model show that gravitational instability drives the non-linear growth of these perturbations, giving rise to the cosmic web: a complex network of voids, filaments, and gravitationally bound structures 
 \citep{peebles_LSS_1980, bond_LSS_1996}.
These virialised objects, known as dark matter haloes, are the fundamental nodes of the cosmic web, acting as the gravitational potential wells where baryonic matter accumulates to form galaxies, groups, and clusters \citep[e.g.,][]{White1978, White1991}. The identification of these haloes in numerical simulations is therefore a central step in connecting theoretical predictions to observable galaxy and cluster populations.

A variety of algorithms have been developed to identify dark matter haloes in simulation outputs. The simplest approaches, such as the friends-of-friends \citep[FoF, ][]{Davis1982, Davis1985} and spherical overdensity \citep[SO,][]{, Warren1992, Lacey1994,compaSO_hadzhiyska2022} techniques, 
locate haloes by tracing overdensities in the matter distribution. Grid-based schemes, such as Amiga's Halo Finder  \citep[\texttt{AHF,}][]{ahf_knollmann2009}, define the density field on an adaptive mesh and iteratively remove unbound particles to identify gravitationally bound structures. More sophisticated phase space algorithms such as \texttt{ROCKSTAR}  \citep{Behroozi2013}, exploit both spatial and velocity information to identify haloes and substructures with high precision. These established tools are essential for constructing mock catalogues \citep[e.g.][]{Springel2005, Hernandez-Aguayo2023, Castander2025}, which are indispensable for interpreting galaxy surveys and provide critical input for testing and refining cosmological analysis pipelines. 

Beyond their use in traditional model testing, the landscape of cosmological inference is undergoing a rapid transformation, placing new and stringent requirements on the entire pipeline used to generate synthetic galaxy samples that are compared against observations. Modern approaches, specifically field-level inference 
\citep{Jasche2019, Stadler2023, Nguyen2024, Doeser2024}
and simulation-based inference 
\citep[SBI, e.g.,][]{Ishida2015, Alsing2018, simbig2022, Tucci2024},  require forward models that are extremely fast and designed for massive, high-throughput execution.
Machine learning methods, specially volumetric Convolutional Neural Networks (CNNs), have emerged to address this need.
They have proven effective in cosmology due to their ability to learn hierarchical representations from high-dimensional data. Applications have ranged from density field reconstruction \citep{Chen2025} and direct parameter inference \citep{Pan2020} to modelling non-linear structure formation \citep{flemu_he, flemu_jamieson}.
CNNs can dramatically accelerate the field generation step and are natively GPU-compatible.
However, the step of identifying structures often relies on traditional, iterative, CPU-based halo finders, which become a severe computational bottleneck for these integrated, GPU-accelerated pipelines.

The application of machine learning to resolve such bottlenecks has evolved from simple feature-based classification to deep field-level emulation.
For example, \citet{mlhalo_lucie} used Random Forests to classify dark matter particles into halo/non-halo categories, demonstrating that local density information in the initial conditions is the primary driver of halo collapse, rendering tidal shear features largely redundant.
The introduction of volumetric CNNs enabled the direct mapping of initial density fields to final halo catalogues.
\citet{mlhalo_berger_stein} employed a three-dimensional VNet architecture to segment proto-halo regions in the initial conditions (Lagrangian patches), while \citet{mlhalo_bernardini} extended this approach by decoupling the task into classification and regression networks to better capture the scatter in the halo-mass relation. A distinct line of work treats this problem as a macroscopic generative task, producing mock halo catalogues from coarse density and velocity fields using auto-regressive normalising flows trained against high-resolution N-body references \citep{Pandey_charm2025}. This task has also been addressed using differentiable frameworks that solve the inverse problem, namely, reconstructing the initial conditions \citep{mlhalo_modi,mlhalo_charnock}. A complementary line of work targets the same inverse problem in the more realistic regime where the structure formation or halo identification step is non-differentiable, using learned-optimiser schemes that keep full physics simulations in the loop \citep{doeser2025}. The state-of-the-art has further expanded to include fully differentiable halo mass function predictions that propagate gradients to cosmological parameters \citep{Buisman2025a} and architectures that incorporate velocity field data to resolve non-linearities in high-variance regimes \citep{mlhalo_razavi}.

While the aforementioned methods focus largely on the initial conditions or on approximate predictions, our goal is to explore the feasibility of detecting virialised haloes directly from the fully evolved, non-linear particle distribution, capturing the full complexity of gravitational collapse.
Inspired by the deep learning strategies of \citet{mlhalo_berger_stein} and the architectural design of \citet{flemu_he}, we employ a GPU-native VNet architecture to perform binary classification at the particle level.
The network uses particle positions and velocities as input to assign a halo-membership probability to every simulation particle.
A highly optimised and parallelised, CPU-based FoF algorithm is then applied exclusively to the predicted halo members in order to partition them into individual objects.
This hybrid ``\cnnfof{}''  pipeline offers a pathway to integrate halo identification directly into GPU-accelerated emulation frameworks.
While the time required to solve gravitational dynamics dominates standard N-body simulations, identifying structures becomes the severe computational bottleneck once non-linear density fields are generated via fast field-level emulators. By accelerating the halo-finding stage specifically, our framework addresses this downstream bottleneck. Using the CNN to prune the search space and identify candidate halo members on the GPU, the subsequent clustering task is significantly reduced, since the bulk of the particle classification is performed in a single forward pass.
This allows the hybrid pipeline to remain compatible with the massive throughput requirements of next-generation inference techniques. 

This paper is organised as follows.
In Section~\ref{sec:dateset_generation}, we describe the methodology used to generate the training samples from cosmological N-body simulations, including the extraction of particle features and the labelling procedure required for supervised learning.
Section~\ref{sec:classification} introduces the architecture employed for binary classification, designed to predict whether individual dark matter particles belong to a halo or not. We also outline the training procedure and optimisation strategy adopted to achieve an accurate classification.
Section~\ref{sec:halo_centres} presents our approach to grouping the filtered particles into distinct objects and recovering halo centres, based on the application of a highly optimised and parallelised FoF clustering algorithm to the particles classified as halo members.
Using this method, we reconstruct halo catalogues and evaluate their consistency with the expected halo distributions.
In Section~\ref{sec:classification_analysis}, we analyse the performance of the complete pipeline.
We begin by evaluating the CNN model and assessing its predictive accuracy.
We then examine the recovered halo properties, with particular emphasis on the halo mass function, complemented by an internal structural analysis of the catalogue objects, in order to assess their agreement with the testing sample.
Finally, Section~\ref{sec:conclusion} summarises our findings and outlines potential directions for future improvements.

\section{Simulation data set}
\label{sec:dateset_generation}

In this Section, we describe the generation and preparation of the cosmological N-body simulation data used for model training and analysis, including the procedure for assigning binary labels to individual dark matter particles.

We generated a suite of cosmological N-body simulations using the publicly available code \texttt{GADGET-4} \citep{gadget4}, assuming a flat $\Lambda$CDM cosmology  at redshift $z=0$. The cosmological parameters used for all simulations are listed in Table \ref{tab:cosmology_table}. The initial conditions were generated at a starting redshift of $z_\mathrm{IC}=99$ from a uniform cubic grid using the \textsc{2LPTic} code \citep{Crocce2006, Crocce2012}.

\begin{table}
    \centering
    \caption{
    Cosmological parameters of the N-body simulations used in our analysis. We assume a flat $\Lambda$CDM cosmological model. We list the physical baryon and cold dark matter densities, $\omega_{\rm b}$, $\omega_{\rm c}$, the dimensionless Hubble parameter, $h$, and the amplitude and spectral index of the primordial scalar mode, $A_{\rm s}$ and $n_{\rm s}$.
    }
    \label{tab:cosmology_table}
    \begin{tabular}{lc}
        \hline
        \textbf{Parameter} & \textbf{Value} \\
        \hline
        $\omega_{\rm b}$ & $0.02244$ \\
        $\omega_{\rm c}$ & $0.1206$ \\
        $h$ & $0.67$ \\
        $n_{\rm s}$ & $0.96$ \\
        $A_{\rm s}$ & $2.1\times 10^{-9}$ \\
        \hline
    \end{tabular}
\end{table}

We considered four simulation configurations to span a range of box sizes and resolutions: $L200$-$N32^3$, $L200$-$N64^3$, $L200$-$N128^3$, and a high-resolution configuration, $L100$-$N128^3$\footnote{The actual box size is $93.75~\mathrm{Mpc}$, chosen to match the resolution of the AletheiaEmu simulation suite of \citet{Sanchez2026} and \cite{FiorilliHMF2026}.}. Here, $L$ denotes the simulation box length in ${\rm Mpc}$ and $N$ refers to the total number of dark matter particles. Higher resolutions provide finer spatial sampling of the non-linear matter field, enabling better characterisation of the internal structure of dark matter haloes. The resulting mass per dark matter particle, $m_{\rm p}$, for the different configurations is summarised in Table \ref{tab:mass_table}.
Unless otherwise stated, we refer throughout this work to the highest resolution configuration, as all resolutions exhibit similar results.

\begin{table}
    \centering
    \caption{
    Dark matter particle mass, $m_{\rm p}$, for the different resolution configurations used in this work. Each configuration is specified by its box size $L$ in ${\rm Mpc}$, and the total number of dark matter particles, $N$.
    }
    \label{tab:mass_table}
    \begin{tabular}{lc}
        \hline
        \textbf{Configuration} & \textbf{$m_{\rm p}$ [$M_\odot$]} \\
        \hline
        $L200-N32^3$  & $9.6921 \times 10^{12}$ \\
        $L200-N64^3$  & $1.2115 \times 10^{12}$ \\
        $L200-N128^3$ & $1.5144 \times 10^{11}$ \\
        $L100-N128^3$ & $1.7681 \times 10^{10}$ \\
        \hline
    \end{tabular}
\end{table}

We identified gravitationally bound structures in these simulations using the \texttt{ROCKSTAR} halo finder \citep{Behroozi2013}, which is recognised for its high accuracy in phase space. The haloes were defined using two different overdensity-based mass definitions: $M_{\mathrm{vir}}$ and $M_{200\mathrm{b}}$. The subscript 'b' indicates multiples of the comoving background matter density, $\rho_\mathrm{b}$, while the virial mass overdensity threshold was defined by the fitting formula given by \cite{bryan_norman_1998} for flat cosmologies. These definitions corresponded to the radii $r_{\mathrm{vir}}$ and $r_{200\mathrm{b}}$, respectively, which defined the spherical boundary of the halo.

To generate the ground-truth labels for our supervised learning task, a particle was labelled as a halo member ($y=1$) if it was identified as gravitationally bound and lay within the overdensity radius of any halo found by \texttt{ROCKSTAR}.
All remaining particles were labelled as non-halo particles ($y=0$).
We restricted our final halo catalogues to objects containing 25 or more member particles to ensure that the identified haloes are numerically stable, avoiding poorly resolved transient overdensities dominated by sampling noise.

For each resolution, the full simulation suite was randomly divided into three subsets for training, validation, and testing. The training set comprised 350 simulations, providing a broad sampling of structural diversity. The validation set contained 50 simulations, employed to monitor convergence and prevent overfitting, while the test set included 100 simulations and served as an independent benchmark for performance assessment. Each resolution thus had its own triplet of datasets, maintaining a uniform and reproducible framework for model comparison.

\section{Hybrid Halo-Finding Pipeline}
\label{sec:methodology}

\subsection{CNN for binary classification}
\label{sec:classification}

Our particle classification network is based on the three-dimensional version of the VNet architecture implemented in the $\mathrm{D^3M}$ framework of \citet{flemu_he}, which was developed to reproduce the non-linear evolution of cosmological density fields in N-body simulations. The $\mathrm{D^3M}$ model extends the classical UNet design of \citep{ronneberger2015unet} to three dimensions and operates intrinsically at the particle level, learning a one-to-one mapping between input and output voxels corresponding to individual simulation particles. This formulation preserves the physical correspondence between each particle and its predicted quantities.

We adapted the $\mathrm{D^3M}$ framework to perform voxel-wise binary classification, where each voxel retains its correspondence with a single simulation particle. In this formulation, the network infers, for every particle, the probability of being a halo member. The architecture follows a standard encoder–decoder layout with skip connections, allowing simultaneous recovery of large-scale structural context and fine-scale spatial features that are essential for accurate halo identification. The details of the final configuration are shown in Table~\ref{tab:cnn_architecture_summary}.

\begin{table}
\centering
\caption{Summary of our CNN architecture for particle classification.}
\label{tab:cnn_architecture_summary}
\begin{tabular}{lc}
\hline
\textbf{Parameter} & \textbf{Value} \\
\hline
Input Channels & 6 (displacement + velocity) \\
Output Channels & 1 (halo probability) \\
Architecture Depth & 3 downsampling + 3 upsampling levels \\
Conv3D Layers per Level & 2 \\
Initial Filters & 64 \\
Maximum Filters & 256 \\
Activation Functions & ReLU (intermediate), Sigmoid (final) \\
Loss Function & BCE \\
Optimizer & Adam (lr = 0.001) \\
Input Shape & $N_{\mathrm{res}} \times N_{\mathrm{res}} \times N_{\mathrm{res}} \times 6$ \\
Output Shape & $N_{\mathrm{res}} \times N_{\mathrm{res}} \times N_{\mathrm{res}} \times 1$ \\
Trainable Parameters & $\sim 8.4 \times 10^6$ \\
\hline
\end{tabular}
\end{table}

The encoder (contracting path) comprises three downsampling stages, each containing two $3 \times 3 \times 3$ convolutional layers with stride~1 and periodic padding, which preserve spatial dimensions while increasing the number of feature maps, followed by a stride-2 convolution that performs spatial downsampling. The number of channels doubles at each level, from 64 to 128 and then 256, enabling the model to capture progressively coarser and more complex representations of the density field. The decoder (expanding path) mirrors this structure, using transposed convolutions for upsampling followed by two standard convolutional layers at each level. Skip connections concatenate feature maps from the encoder to their corresponding decoder layers, ensuring that fine spatial details are retained and that gradients flow efficiently during training. All intermediate layers employ ReLU activations and batch normalisation to enhance convergence stability and mitigate internal covariate shift.

To adapt the architecture for binary classification, the final layer of the decoder is modified to produce a single output channel per voxel, followed by a sigmoid activation function:
\begin{equation}
    \phi(x) = \frac{1}{1 + e^{-x}}.
\end{equation}
This non-linear function maps each raw output $x$ to a probability in the range $[0,1]$, representing the likelihood that a given particle belongs to a dark matter halo. While the sigmoid activation is widely used in binary classification due to its probabilistic output, it is also susceptible to vanishing gradients in very deep networks. Nonetheless, for our architecture depth, it provides a suitable mechanism for assigning soft binary labels.

The initial Lagrangian particle distribution is embedded into a regular three-dimensional lattice in which each voxel contains exactly one simulation particle. The input to the network is therefore represented as a six-channel voxelised tensor that encodes both spatial and dynamical information for each simulation particle. Each voxel contains the three Cartesian components of the displacement field, $\boldsymbol{\Psi} = (\Psi_x, \Psi_y, \Psi_z)$, representing deviations from the initial grid positions, together with the three components of the velocity vector, $\boldsymbol{v} = (v_x, v_y, v_z)$, describing the particle velocities.

The complete input volume is therefore structured as
\begin{equation}
X \in \mathbb{R}^{N_{\mathrm{res}} \times N_{\mathrm{res}} \times N_{\mathrm{res}} \times 6},
\end{equation}
where $N_{\mathrm{res}}$ denotes the resolution of the simulation along one dimension (i.e., $N=N_{\mathrm{res}}^3$ represents the total number of particles). This ensures a one-to-one correspondence between the input of the network and the individual simulation particles. This combination of displacement and velocity information allows the network to effectively infer the signature of gravitationally bound structures from both the positional and dynamical configuration of the matter field.

The final output of the network is a single-channel tensor of the form
\begin{equation}
Y \in \mathbb{R}^{N_{\mathrm{res}} \times N_{\mathrm{res}} \times N_{\mathrm{res}} \times 1},
\end{equation}
in which each voxel contains a scalar probability score that can be thresholded or interpreted directly to determine halo membership.

The model was trained in a supervised fashion using ground-truth labels derived from the \texttt{ROCKSTAR} halo catalogue as described in Section~\ref{sec:dateset_generation}. The network was optimised using the binary cross-entropy (BCE) loss function,
\begin{equation}
\mathcal{L}_\mathrm{BCE} = - \left[ y \log(\hat{y}) + (1 - y) \log(1 - \hat{y}) \right],
\label{eq:loss}
\end{equation}
where $y \in \{0,1\}$ denotes the ground-truth label and $\hat{y}$ is the predicted probability from the network output. This formulation is standard for binary classification tasks with probabilistic predictions \citep{ML2:01}. Optimisation was performed using the Adam algorithm \citep{kingma_adam} with default parameters and a fixed learning rate of $10^{-3}$. Random shuffling was applied across mini-batches of 3D volumes to improve generalisation, and early stopping based on validation loss was employed to prevent overfitting.

We set a maximum training duration of 120 epochs, determined empirically to ensure full convergence while maintaining computational efficiency. Additional runs confirmed that the validation loss consistently reached its minimum before this limit (typically between 50 and 110 epochs), indicating that convergence was achieved well before the upper bound. The model weights corresponding to the epoch with the minimum validation loss were stored, providing an empirical estimate of the best-performing model for each resolution.

The full training pipeline was implemented in \texttt{Python} using the \texttt{PyTorch} deep-learning framework \citep{PyTorch_Paszke}. Multi-GPU parallelisation was achieved through the \texttt{nn.DataParallel} module, which enabled the use of large batch sizes and substantially reduced training time. All models were trained on the high-performance Raven GPU cluster at the Max Planck Computing and Data Facility (MPCDF\footnote{\url{https://www.mpcdf.mpg.de}}), taking advantage of the high throughput of NVIDIA A100 GPUs for tensor operations.

\subsection{FoF algorithm to group distinct haloes}
\label{sec:halo_centres}

As described in the previous section, the CNN step identifies particles that belong to haloes. 
However, it does not associate these particles with individual halo clumps. For the particular redshift and cosmology considered here, this classification step dramatically reduces the set of particles that must be further considered to roughly $38\%$ of the total number of particles. While this percentage is specific to the $L100$-$N128^3$ simulation and varies slightly with resolution, the resulting data is inherently clumpy, as the predicted halo particles occupy localised regions in space separated by large empty areas. This structure is the key element that enables the optimisation of the subsequent halo-finding step.

To identify the individual haloes and their particle memberships, we run a classic FoF algorithm, which computes pairwise particle distances and assigns the same halo membership to particles whose separation is below a given linking length, $\lambda$. This threshold length is typically defined as a small fraction of the mean interparticle separation. 
In our case, we adopt $\lambda = b\,\bar{n}^{-1/3}$, where $\bar{n}$ is the mean number density of the full original particle distribution, and $b=0.25$. This value is validated through a systematic calibration presented in Appendix~\ref{subsec:linking_length}. The grouping process is transitive, resembling a breadth-first search (BFS) on a graph where nodes are particles connected by edges shorter than the linking length.

Since traditional FoF algorithms based on direct particle searches are inherently serial, their efficient parallelisation on concurrent computing systems is complex. Our FoF approach exploits two facts: first, the input data are already highly clustered (thanks to the preceding CNN step) and naturally separated into distinct spatial domains. Second, we reduce the cost of pairwise distance calculations by implementing a grid technique.

For the second point, we divide the simulation box into regular cubic grids, or voxels, and reorder the input particle array along a Peano-Hilbert space-filling curve. This optimisation improves cache behaviour in the memory-demanding stages of computation. The voxel methods used here have proved highly efficient in other contexts, such as void finding algorithms, and we adopt the publicly available functions for voxel manipulation from the \texttt{Popcorn} void-finder code \citep{Popcorn2023}.

We run a two-stage FoF search. First, a voxel-level FoF search is performed. At each BFS stage, we identify, in parallel, every group of non-empty voxels whose minimum separation (the distance between the closest vertices) is below the linking length. Parallel efficiency is limited here by the need to prevent race conditions, forcing some execution parts to remain serial. The cost of this stage scales with the number of non-empty voxels rather than the total particle count. Second, after isolating these connected voxel groups, we carry out a particle-level FoF run, assigning each spatial domain to its own thread. This decomposition makes the workload embarrassingly parallel, eliminates race conditions in halo assignment, and allows us to significantly outperform traditional particle-based FoF finders. A final optimisation from the \texttt{Popcorn} library links particles in voxel pairs whose maximum separation is below the linking length in one step, removing many needless pairwise distance calculations.

Individual halo structures are identified by applying the optimised FoF clustering algorithm to the set of particles predicted as halo members by the CNN.
This procedure groups the majority of particles into distinct haloes, for which we define the spatial centre as the centre of mass of the constituent particles.
To ensure that the final catalogue is complete, any remaining particles—classified as halo members by the CNN but not linked to a group by the FoF—are subsequently assigned to their nearest identified halo centre.
The total mass of each halo is then computed as the sum of the masses of its FoF-linked particles plus any assigned residual particles.
This approach leverages the FoF algorithm to locate and deblend structures while ensuring every particle detected by the neural network contributes to the final mass budget.

\subsection{Motivation for our pipeline design}
\label{subsec:pipeline_motivation}

This section serves to clarify the motivation behind our pipeline design by addressing three aspects: the positioning of our method with respect to other approaches in the literature, the necessity of a modular two-stage architecture (CNN and FoF), and the ultimate goal we aim to achieve by maintaining particle-level fidelity within future cosmological pipelines.

As outlined in the Introduction, modern machine learning frameworks frequently approach structure identification by replacing the discrete N-body particle distribution with grid-based density approximations or generative models.
While these methodologies are effective at their respective goals, such as predicting statistical halo properties or reconstructing initial conditions, they bypass exact, high-resolution particle dynamics.
Consequently, they are not designed to provide a physical link between a generated halo and the specific underlying dark matter particles.

In contrast, our CNN+FoF pipeline is fundamentally a classifier, not a generator.
Our aim is to emulate a phase-space halo finder.
Operating on the fully evolved, non-linear particle distribution, our network classifies actual simulation particles.
In this way, we try to explicitly preserve the phase-space history of haloes, such as their mass, centre of mass, velocity and internal substructure, features that are often lost or merely statistically sampled in other approaches.

Understanding our approach as a particle classifier helps to understand the specific roles within our modular architecture.
On the one hand, the CNN acts as a binary filter to separate halo from non-halo particles.
The primary contribution of the CNN is twofold: massive data reduction (approximately $62\%$ of the particles are discarded) and the injection of phase-space physical fidelity into the selection.
However, it does not, by itself, distinguish between different clumps or build a catalogue.
That is the role of the FoF stage.
Its purpose is just to label or give an identity to different objects in order to obtain a distinctive catalogue of haloes.

As noted in the literature \citep[e.g.][]{mlhalo_berger_stein}, many hybrid pipelines require post-processing to identify individual objects, but the time performance of these steps is often overlooked.
One of the goals of this work was to maintain overall execution times faster than the original baseline method.
We experimented with various clustering algorithms, including DBSCAN \citep{dbscan}, but FoF proved to be the simplest and fastest for this specific task.

It is worth highlighting that the FoF algorithm used in our pipeline is very different from the standard vanilla implementation.
It has been efficiently parallelised, as described in Section~\ref{sec:halo_centres}.
This parallelisation is only possible on a highly clumpy particle distribution.
In other words, this FoF implementation depends strongly on the previous filtering step.

Regarding the overall computational acceleration aspect, our vision is to integrate our CNN+FoF method with future, more general, ultra-fast, end-to-end cosmological pipelines.
In particular, we aim to pair this tool with modern field-level emulators \citep[e.g.][]{flemu_he, flemu_jamieson} that correct the positions and velocities of cheap approximate simulation particles to generate fully evolved N-body-like distributions.
These emulators bypass the traditional N-body simulation generation entirely, accelerating the process by several orders of magnitude.
Moreover, this ensures the whole process preserves the physical link between matter particles and haloes.
Once the simulation bottleneck is removed, complex phase-space structure identification becomes the new primary bottleneck.
Our pipeline is designed precisely to tackle this challenge.

Ultimately, our method represents a proof of concept that demonstrates it is possible to replicate the complex logic of a phase-space halo finder using deep learning, while simultaneously maintaining high physical fidelity.
While traditional phase-space finders are themselves undergoing significant computational optimisations, such as the recent MPI-parallel implementation of \texttt{ROCKSTAR} \citep{Tokuue_mpi_rockstar2026}, our deep learning framework provides a complementary pathway natively suited for integration with modern AI-accelerated pipelines.
We recognise our current modularity is a first step, and we intend in future work to enhance our pipeline by transitioning the FoF stage into a fully GPU-integrated step.
This will be explored either by leveraging a recently developed GPU-native FoF algorithm \citep{Horowitz_jfof2025} or by absorbing the clustering logic directly into the CNN architecture itself.

\section{Results}
\label{sec:classification_analysis}

In this section, we present the results of applying the hybrid \cnnfof{} pipeline to the N-body simulation test set. We first focus on assessing the particle-level classification accuracy, and later on properties of the recovered haloes, comparing the statistical and structural properties of the final halo catalogue to the \texttt{ROCKSTAR} reference. We also compare the computational performance of our pipeline against a full \texttt{ROCKSTAR} run. While the results in this section adopt $M_{\rm 200b}$ as the primary reference, the corresponding analysis for $M_{\rm vir}$ is provided in Appendix~\ref{app:mvir}.
Moreover, we focus the analysis mainly on the $L100$-$N128^3$ configuration, with the other cases leading to similar results.

\subsection{Particle classification accuracy}
\label{subsec:cnn_validation}

To quantify the accuracy of the classification performed by our network into halo and non-halo particles, we categorise each particle-label prediction based on its agreement with the ground-truth labels from \texttt{ROCKSTAR}.
Particles identified as halo members by both \texttt{ROCKSTAR} and the classification network are termed true positives (TP), while those classified as halo particles by the CNN but not by \texttt{ROCKSTAR} are defined as false positives (FP). Conversely, false negatives (FN) correspond to particles that \texttt{ROCKSTAR} labels as halo members but the CNN predicts as non-halo. Finally, true negatives (TN) denote the particles consistently recognised as non-halo by both 
methods.

\begin{figure}
    \centering
    \includegraphics[width=\columnwidth, height=0.9\columnwidth]{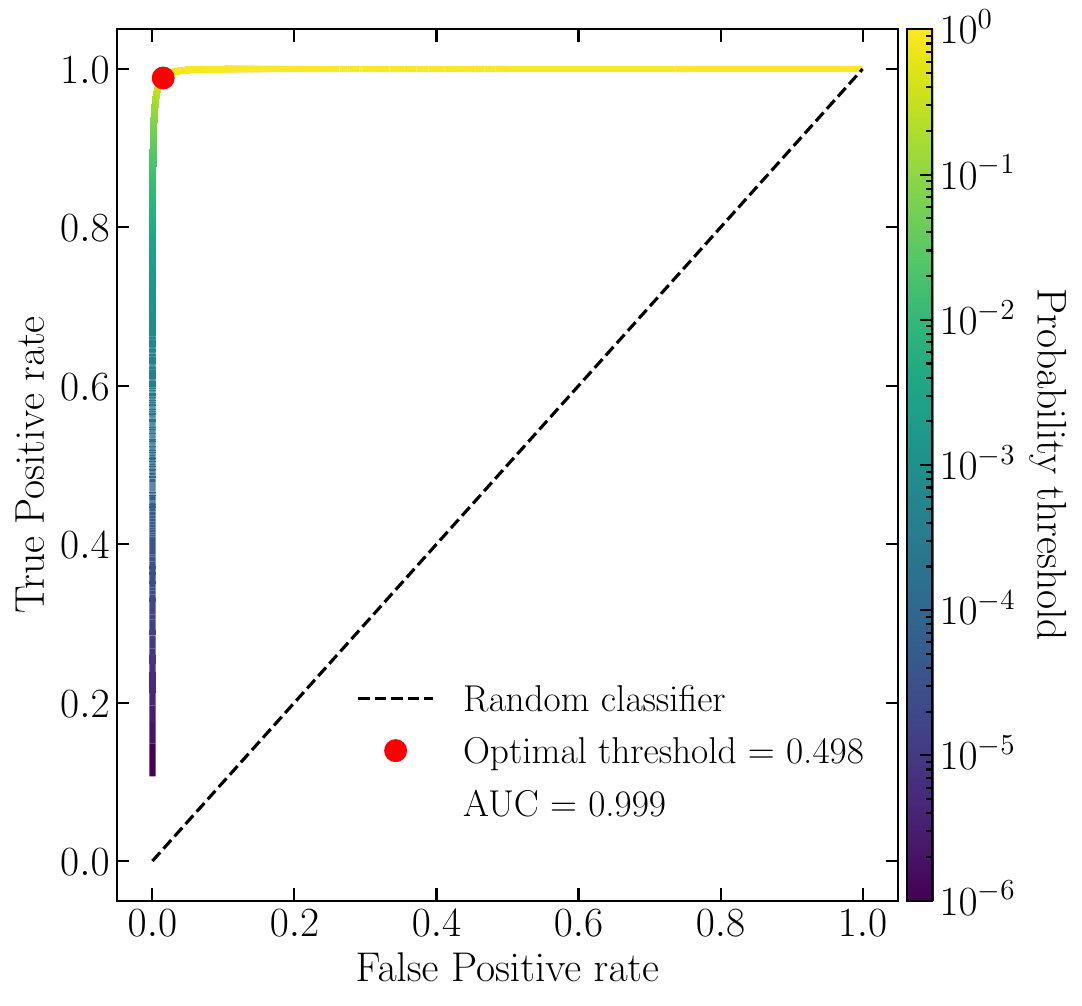}
    \caption[ROC Curve for classification network]{ROC curve showing the CNN's ability to distinguish halo and non-halo particles across varying probability thresholds. The red point indicates the best threshold for the classification, corresponding to $y=0.498$.}
    \label{fig:roc_curve}
\end{figure}

Figure~\ref{fig:roc_curve} presents the receiver operating characteristic (ROC) curve corresponding to the high-resolution $L100$-$N128^3$ configuration, which traces the trade-off between the true positive rate and the false positive rate across varying classification thresholds. The red point marks the optimal threshold of $y=0.498$, which we used for binary classification throughout our analysis.

Using this optimal threshold, the predicted probabilities are converted into binary labels and compared against the ground-truth labels. The resulting confusion matrix, detailing the fraction of particles in each category, is shown in Table~\ref{tab:conf_mat}.

\begin{table}
    \centering
    \caption{
    Confusion matrix of the CNN particle classification for the high-resolution $L100-N128^3$ configuration. Values represent the percentage of the total particle population falling into each category: true positives (TP), false positives (FP), true negatives (TN), and false negatives (FN).
    }
    \label{tab:conf_mat}
    \renewcommand{\arraystretch}{1.3} 
    \setlength{\tabcolsep}{12pt}      
    \begin{tabular}{l cc}
        \hline
         & \multicolumn{2}{c}{\textbf{Predicted Class}} \\
        \cline{2-3} 
        \textbf{True Class} & \textbf{Halo} & \textbf{Non-Halo} \\
        \hline
        \textbf{Halo}      & $36.25\%$ (TP) & $0.71\%$ (FN) \\
        \textbf{Non-Halo}  & $0.60\%$ (FP)  & $62.44\%$ (TN) \\
        \hline
    \end{tabular}
\end{table}

The classification performance of the trained CNN is further evaluated using standard metrics computed on the test simulations, which are reported in Table~\ref{tab:metrics_table} for all resolution configurations. Definitions and notation for all metrics are provided in Appendix~\ref{app:metrics}. 
For the highest resolution, the network attains an accuracy of 98.69\%, with precision and recall values of 98.01\% and 98.42\%, and a specificity of 98.80\%. These results demonstrate that the classifier effectively captures the primary halo features, maintaining a stable balance between precision and recall across the tested resolutions.

\begin{table*}
    \centering
    \caption{
    Performance metrics of the CNN particle classification for all the resolution configurations used in this work. The definitions of the metrics are provided in Appendix~\ref{app:metrics}.
    }
    \label{tab:metrics_table}
    \renewcommand{\arraystretch}{1.3} 
    \setlength{\tabcolsep}{10pt}      
    \begin{tabular}{l cccc}
        \hline
        \textbf{Metric} & \textbf{$L200-N32^3$} & \textbf{$L200-N64^3$} & \textbf{$L200-N128^3$} & \textbf{$L100-N128^3$} \\
        \hline
        Accuracy    & $0.99$ & $0.98$ & $0.98$ & $0.99$ \\
        Precision   & $0.96$ & $0.95$ & $0.97$ & $0.98$ \\
        Recall      & $0.96$ & $0.96$ & $0.97$ & $0.98$ \\
        Specificity & $0.99$ & $0.98$ & $0.99$ & $0.99$ \\
        F1 Score    & $0.96$ & $0.95$ & $0.97$ & $0.98$ \\
        MCC         & $0.98$ & $0.97$ & $0.98$ & $0.98$ \\
        AUC         & $0.98$ & $0.98$ & $0.98$ & $0.99$ \\
        AP          & $0.93$ & $0.95$ & $0.95$ & $0.99$ \\
        \hline
    \end{tabular}
\end{table*}

\begin{figure}
    \centering
    \includegraphics[width=\columnwidth]{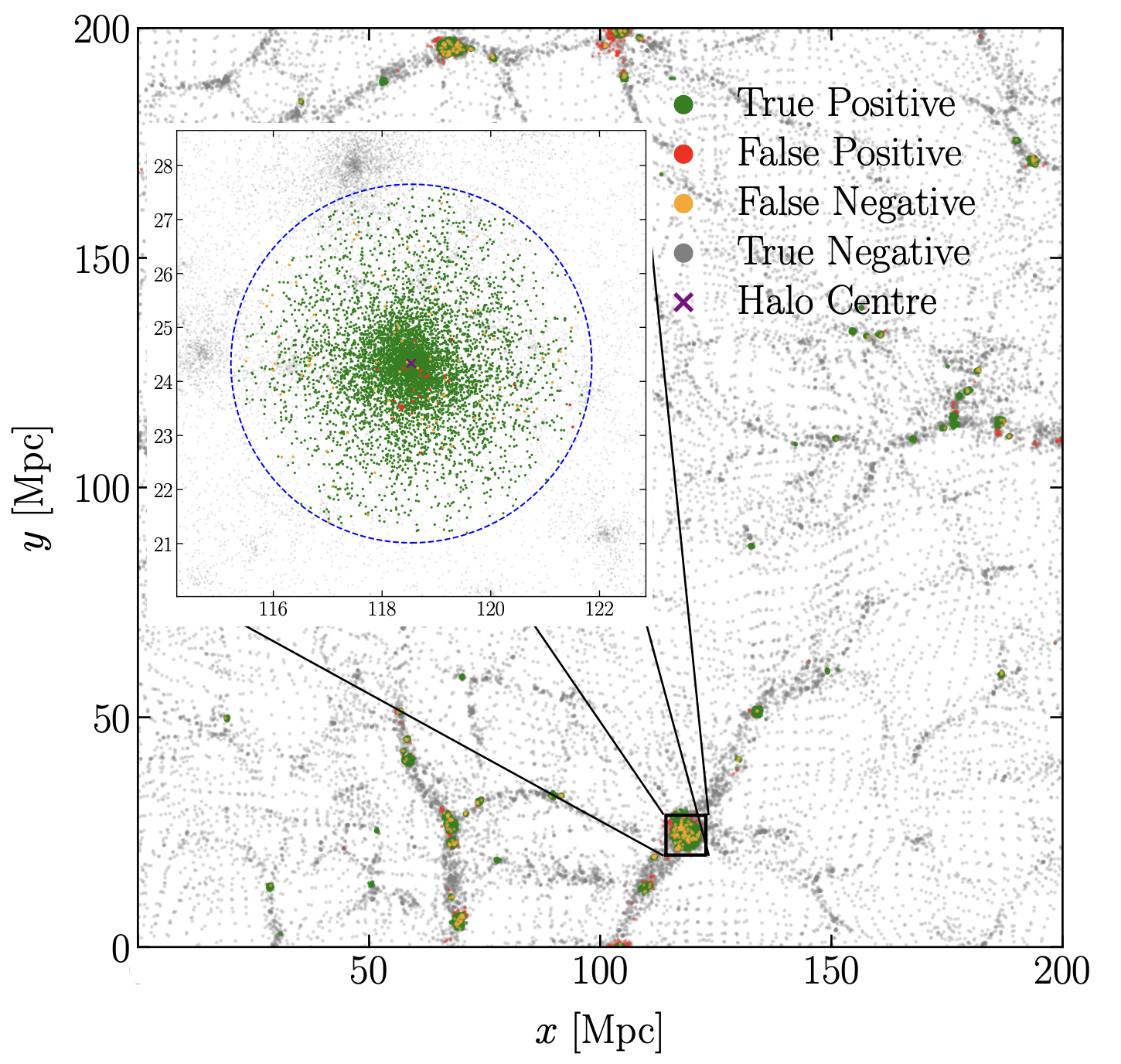}
    \caption[Spatial distribution of classified particles]{
    Spatial distribution of particles in one of the $L200$-$N128^3$ test simulations, colour-coded by classification category: true positives (green), false positives (red), false negatives (orange), and true negatives (grey).
    The main panel displays a projected slice (depth of $2.5\%$ of the box size) illustrating the large-scale cosmic web.
    The inset zooms in on a representative halo identified by \texttt{ROCKSTAR}, with the centre marked by a purple cross and the $r_{200\mathrm{b}}$ radius indicated by a dashed blue circle.
    The network successfully traces the main halo body, while misclassifications are concentrated at the halo outskirts.
    }
    \label{fig:classified_particles}
\end{figure}

The spatial distribution of the classified particles provides further insight into the network's predictive behaviour. Figure~\ref{fig:classified_particles} presents a spatial visualisation of the classified particles in a projected slice of one of the $L200$--$N128^3$ simulations. TP particles (green) form compact, high-density clusters that align with gravitationally bound structures. TN particles (grey) dominate the underdense regions, confirming the network's reliability in identifying the background field.
Misclassified particles, FP (red) and FN (orange), are typically concentrated near the peripheries of haloes where density gradients are steep.

\begin{figure} 
    \centering 
    \includegraphics[width=\columnwidth]{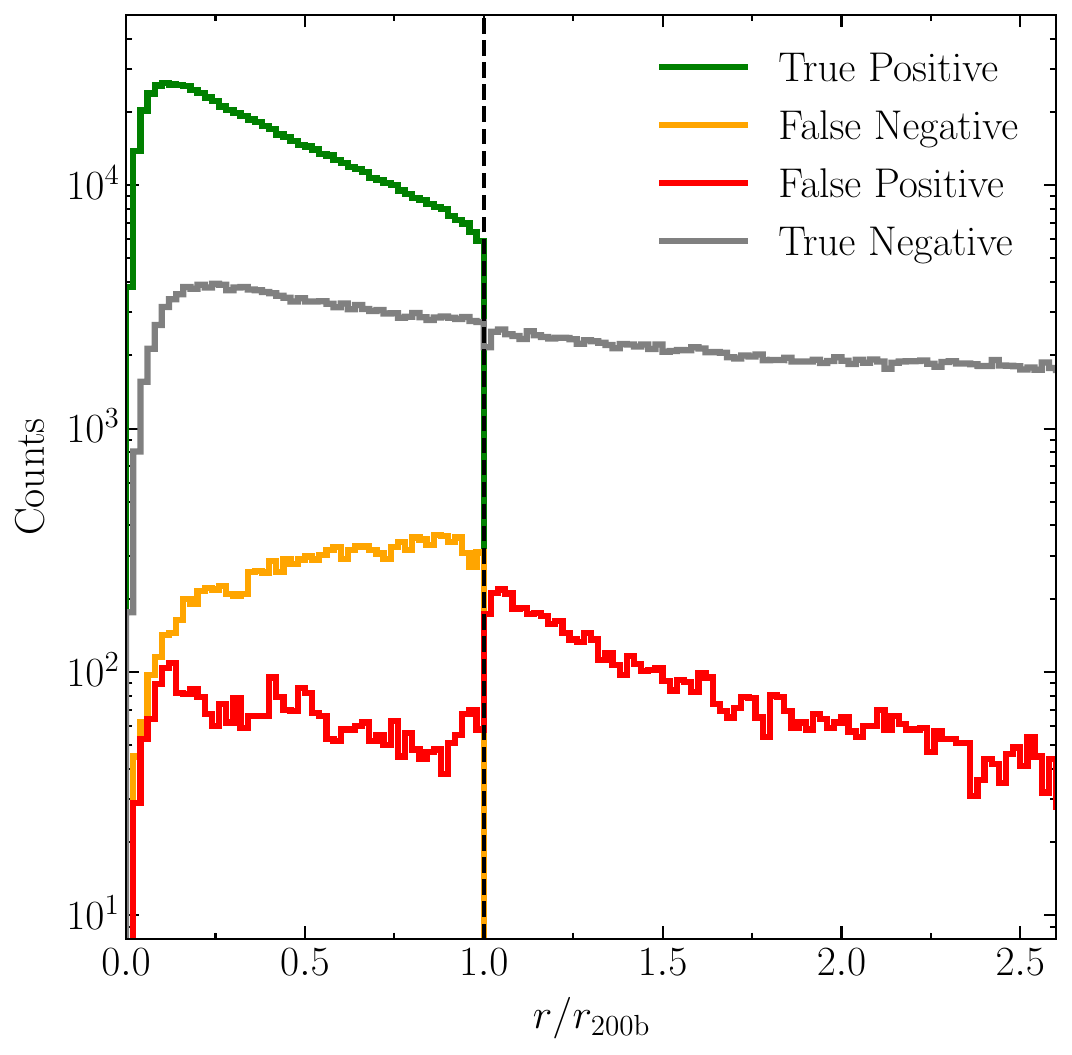} 
    \caption[Normalised radial distance distribution of classified particles]{
    Normalised radial distribution of particles relative to the nearest \texttt{ROCKSTAR} halo centre, scaled by the $r_{200\mathrm{b}}$ radius.
    The vertical dashed line marks the halo boundary at $r = r_{200\mathrm{b}}$.
    Inner regions are dominated by true positives (green), which decline at the halo radius, indicating accurate identification of gravitationally bound particles.
    Misclassifications (false negatives, orange; false positives, red) peak around the halo radius, reflecting the challenge of distinguishing particles located near the boundary.
    True negatives (grey) dominate the outer regions, although there is a significant fraction of them located inside $r_{200\mathrm{b}}$.
    Histogram counts are shown on a logarithmic scale.
    }
    \label{fig:particles_hist}
\end{figure}

To quantify this effect, we computed the normalised radial distance, $r/r_{200\mathrm{b}}$, between each particle and its closest \texttt{ROCKSTAR} halo centre.
Figure~\ref{fig:particles_hist} shows the distribution of the normalised radial distance $r/r_{200\mathrm{b}}$ for particles in the TP, FP, FN, and TN groups.
Notice that the TP and FN distributions are confined within $r/r_{200\mathrm{b}} \leq 1$ by definition.
The TP distribution, on the one hand, shows a strong concentration towards the centre, consistent with the spherical overdensity criterion used to define halo membership and indicating correct identification of particles associated with halo centres.
The FN distribution, on the other hand, peaks near $r/r_{200\mathrm{b}} = 1$, indicating that incorrect halo assignments occur primarily in regions close to the halo boundary where the transition between halo and background particles occurs.
A similar behaviour is observed for FP particles, which are also mostly located near $r/r_{200\mathrm{b}} = 1$, reflecting the difficulty of classification in boundary regions where particle densities and binding properties vary.
Unlike FN, FP particles can go beyond $r/r_{200\mathrm{b}} = 1$, indicating that lower mass clumps were also identified by the CNN, although they do not qualify for \texttt{ROCKSTAR} requirements.
Finally, TN particles follow a broad distribution, consistent with their association with the background field.
Notice also that there are significant TN inside the boundary marked by $r=r_{200\mathrm{b}}$.
These distributions show that the classifier does not rely solely on radial distance but distinguishes halo membership based on learned phase space information.
Overall, the radial distributions show that the network reproduces the spatial structure of haloes, with accurate classification in central regions and consistent behaviour near halo boundaries.

\subsection{Recovering halo properties}
\label{subsec:recovered_halos}

After particle classification, the optimised FoF algorithm was applied to the predicted halo members to build the final halo catalogues.
To evaluate the consistency and precision of the resulting catalogue, we calculated the distance between the centres of the FoF haloes and that of the nearest \texttt{ROCKSTAR} halo, $d_\mathrm{FR}$. A match between both catalogues was defined as the cases in which this distance was less than the \texttt{ROCKSTAR} $r_{200\mathrm{b}}$ radius.
In cases where multiple FoF haloes were found within this radius, only the nearest was retained.

To quantify the fidelity of the \cnnfof{} catalogue relative to the \texttt{ROCKSTAR} reference, we evaluate two primary metrics: purity and completeness.
Purity, $P$, measures the fraction of recovered haloes that are real (i.e., have a counterpart in the reference catalogue), while completeness, $C$, measures the fraction of reference haloes successfully recovered by our pipeline.
Formally, we define these metrics as a function of halo mass, $M$:
\begin{equation}
\label{eq:purity_completeness}
\begin{split}
P(M) &= \frac{N_{\rm matched}(M)}{N_{\rm \cnnfof{}}(M)}\, , \\
C(M) &= \frac{N_{\rm matched}(M)}{N_\texttt{ROCKSTAR}(M)}\, .
\end{split}
\end{equation}

Table~\ref{tab:matching_table} summarises the global values of these statistics (over the full mass range) for all resolution configurations.
For the highest resolution, the pipeline achieves a global completeness of $90.54\%$, indicating that the vast majority of reference haloes are successfully recovered.
The catalogue also exhibits high fidelity, with a global purity of $99.71\%$, corresponding to a low spurious detection rate.

\begin{table*}
    \centering
    \caption{
    Global purity and completeness for the \cnnfof{} halo catalogues across different resolutions. Purity is defined as the fraction of recovered haloes that have a counterpart in the reference \texttt{ROCKSTAR} catalogue, while completeness denotes the fraction of reference haloes successfully recovered by the pipeline.
    }
    \label{tab:matching_table}
    \renewcommand{\arraystretch}{1.3} 
    \setlength{\tabcolsep}{14pt}
    \begin{tabular}{l cccc}
        \hline
        \textbf{Metric} & \textbf{$L200-N32^3$} & \textbf{$L200-N64^3$} & \textbf{$L200-N128^3$} & \textbf{$L100-N128^3$} \\
        \hline
        Purity       & $71.55\%$ & $99.57\%$ & $99.90\%$ & $99.71\%$ \\
        Completeness & $63.48\%$ & $76.75\%$ & $88.45\%$ & $90.54\%$ \\
        \hline
    \end{tabular}
\end{table*}

\begin{figure}
    \centering
    \includegraphics[width=\columnwidth]{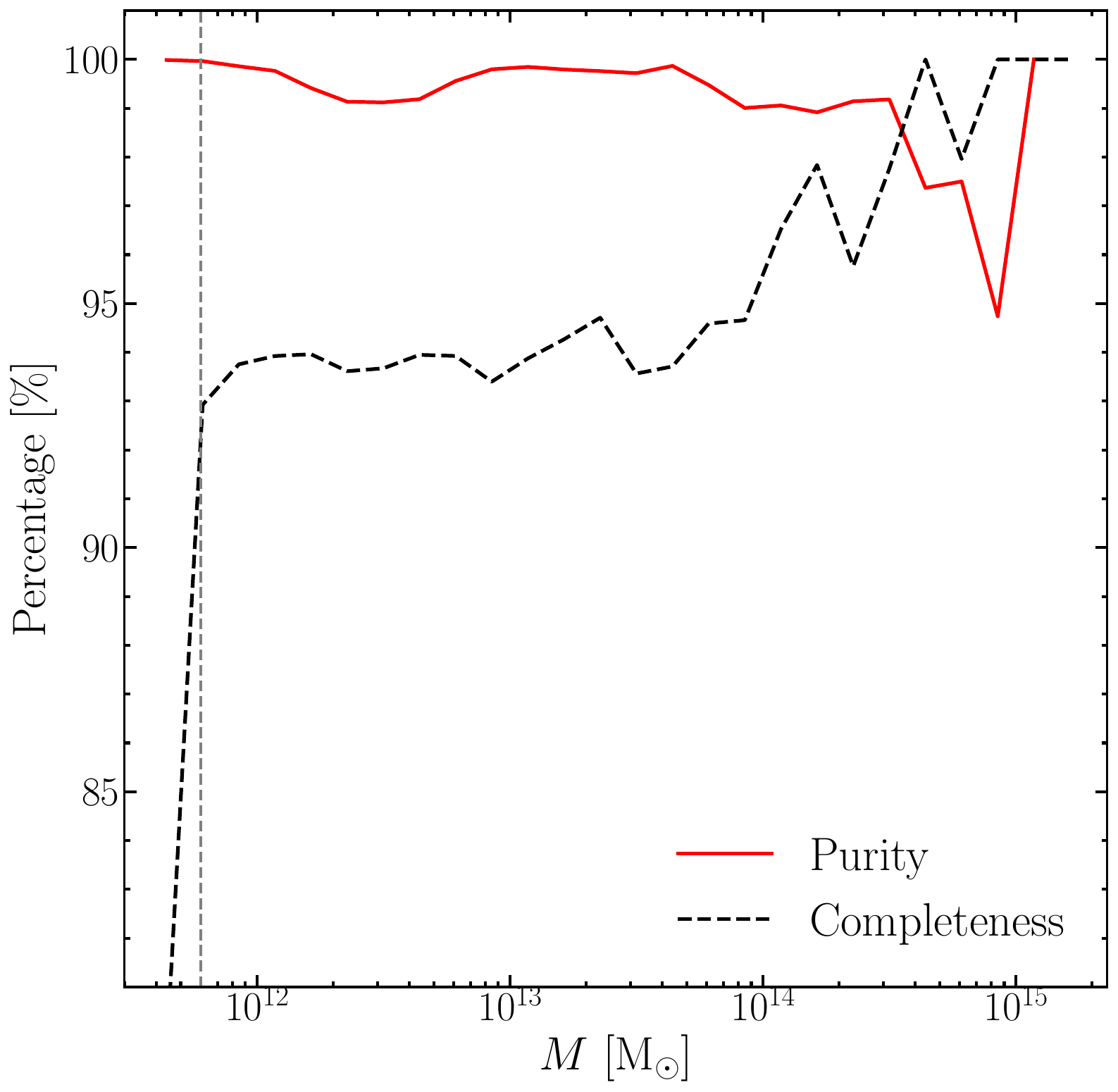}
    \caption[Purity and completeness as a function of halo mass]{
    Mass-dependent purity (solid red) and completeness (dashed black) of the \cnnfof{} halo catalogue relative to the reference \texttt{ROCKSTAR} sample. 
    The completeness exceeds $90\%$ across a wide mass range, remaining stable at $\sim 94\%$ above $M = 5 \times 10^{11} \, M_{\odot}$ (grey vertical line).
    The drop at the low-mass arises as we approach the resolution limit.
    Conversely, the purity generally exceeds $99\%$, but exhibits a mass-dependent decline.
    This drop can be attributed to FoF fragmentation at lower masses and to the merging of adjacent structures at the high-mass end.
    }
    \label{fig:halo_matchings}
\end{figure}

While these global statistics demonstrate robust overall performance, the reconstruction quality varies with halo mass.
Figure~\ref{fig:halo_matchings} shows purity and completeness as functions of halo mass according to the definitions given in equation~(\ref{eq:purity_completeness}).
The purity remains high across a broad mass range, generally exceeding $99\%$, although it exhibits a mass-dependent decline.
We attribute the drop at the low-mass end to fragmentation by the FoF algorithm, which artificially classifies substructures as satellites around massive host haloes, while the decrease at the high-mass end is likely driven by the merging of adjacent structures.
Overall, the reconstructed catalogue is largely representative of the underlying halo population, with minimal spurious detections in the lower- and intermediate-mass regimes.
The completeness is similarly stable; it exceeds $90\%$ across a wide mass range and remains constant at $\sim 94\%$ above $M = 5 \times 10^{11} \, M_{\odot}$.
The expected decline at the low-mass end arises as we approach the resolution limit, where haloes are composed of fewer particles, leading to an increased sensitivity of these poorly sampled structures to boundary definitions.
Based on these stability metrics, we adopt $M = 5 \times 10^{11} \, M_{\odot}$ as the lower mass limit for the remainder of the analyses in this work.

To evaluate the precision of the recovered halo properties, we compared the centre-of-mass (CM) positions and velocities of the matched haloes with those of the \texttt{ROCKSTAR} halo catalogue as a reference. For each matched pair, we computed the component-wise offsets $\Delta \mathbf{X} = \mathbf{X}^{\rm \cnnfof{}} - \mathbf{X}^{\mathrm{\texttt{ROCKSTAR}}}$,
and normalised them by the \texttt{ROCKSTAR} $r_{200\mathrm{b}}$ radius, yielding the dimensionless quantities 
$\Delta X_i / r_{200\mathrm{b}}$, where $i \in \{x,y,z\}$.
Similarly, for the velocities, we evaluated the component-wise ratios  $V_i^{\mathrm{\texttt{ROCKSTAR}}} / V_i^{\rm \cnnfof{}}$.

\begin{figure}
    \centering
    \includegraphics[width=\columnwidth]{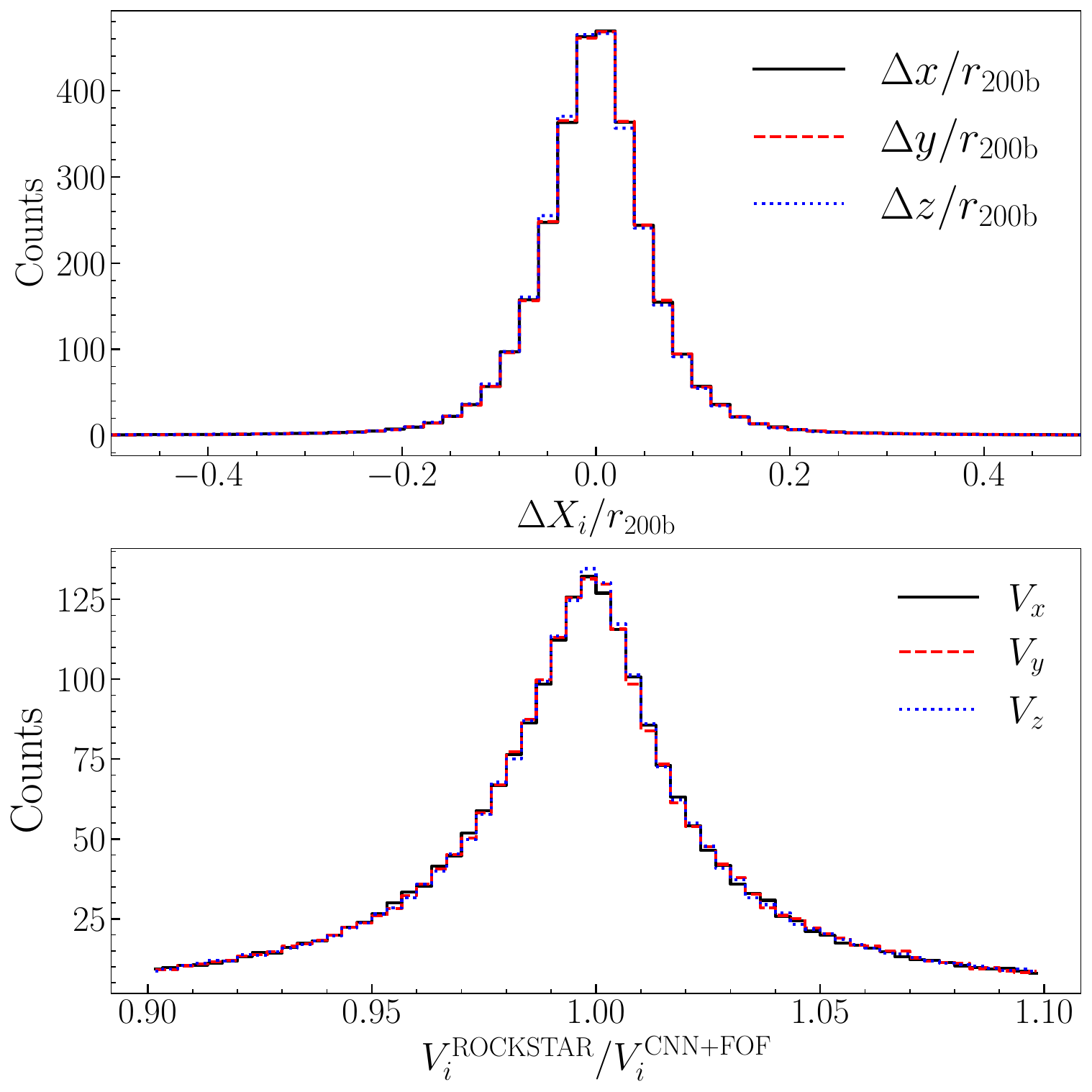}
    \caption[Comparison of recovered positions and velocities]{
    Accuracy of recovered halo properties for the matched \cnnfof{} and \texttt{ROCKSTAR} samples.
    \textit{Top:} Distribution of centre-of-mass position offsets, $\Delta X_i$, normalised by the $r_{200\mathrm{b}}$ radius, along each Cartesian axis.
    \textit{Bottom:} Distribution of the component-wise velocity ratios, $V_i^{\texttt{ROCKSTAR}} / V_i^{\rm \cnnfof{}}$.
    The position offsets are sharply peaked at zero, while the velocity ratios cluster tightly around unity, demonstrating that the pipeline recovers both the spatial and bulk dynamical properties of haloes with high fidelity.
    }
    \label{fig:ratio_pos_vel}
\end{figure}

Figure~\ref{fig:ratio_pos_vel} shows the resulting distributions for the normalised position offsets and velocities. The former are sharply peaked at zero and exhibit symmetric, narrow profiles across all three spatial components, indicating that halo centres are recovered to within a small fraction of the halo radius. The velocity ratios are correspondingly well centred around unity with modest scatter, demonstrating that the reconstructed haloes preserve the bulk kinematic properties of the \texttt{ROCKSTAR} reference catalogue. Together, these results confirm that the \cnnfof{} pipeline recovers both the spatial and dynamical properties of haloes.

\begin{figure}
    \centering
    \includegraphics[width=\columnwidth]{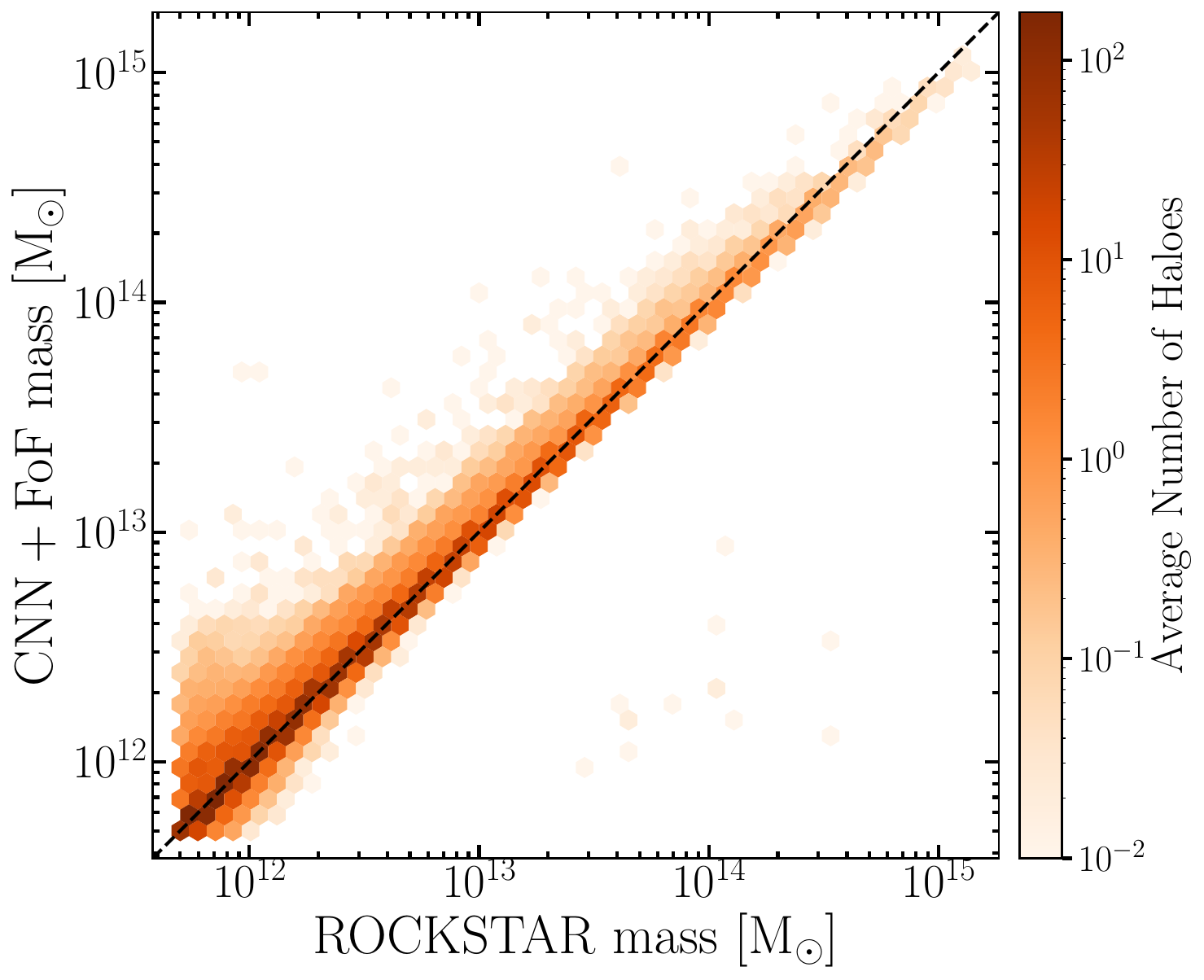}
    \caption[ratio of two masses]{
    Comparison of the halo masses recovered by the hybrid \cnnfof{} pipeline against the reference values from the \texttt{ROCKSTAR} catalogue. The hexagonal bins are colour-coded by the average number of haloes per bin.
    The dashed black line represents the 1:1 relation. The tight distribution along the diagonal across several orders of magnitude demonstrates the high fidelity of the mass recovery, with minimal systematic bias. The scatter at the low-mass end reflects the inherent difficulty in precisely defining boundaries for haloes composed of a limited number of particles. 
    }
    \label{fig:mass-mass}
\end{figure}

To further assess the performance of our pipeline, we compared the total masses of the recovered haloes with their corresponding values in the \texttt{ROCKSTAR} reference catalogue. This comparison is presented in Figure~\ref{fig:mass-mass} as a density plot, which displays the average distribution of halo masses across the $100$ test set simulations.
The results show a tight linear correlation across the entire mass range, indicating that the combined CNN-classification and proximity-based particle assignment effectively captures the total mass of the underlying gravitational structures. 
At the low-mass end, we note a slight increase in the scatter. This is expected, as haloes in this regime consist of relatively few particles, making their mass estimates more sensitive to small variations in the CNN’s particle-level classification and the subsequent deblending by the FoF stage. However, even in this limit, the pipeline maintains high fidelity, as evidenced by the high purity and completeness metrics discussed previously. For higher-mass systems, the scatter reduces significantly, demonstrating the robustness of the method for the massive clusters that drive many large-scale structure statistics.

\begin{figure}
    \centering
    \includegraphics[width=\columnwidth]{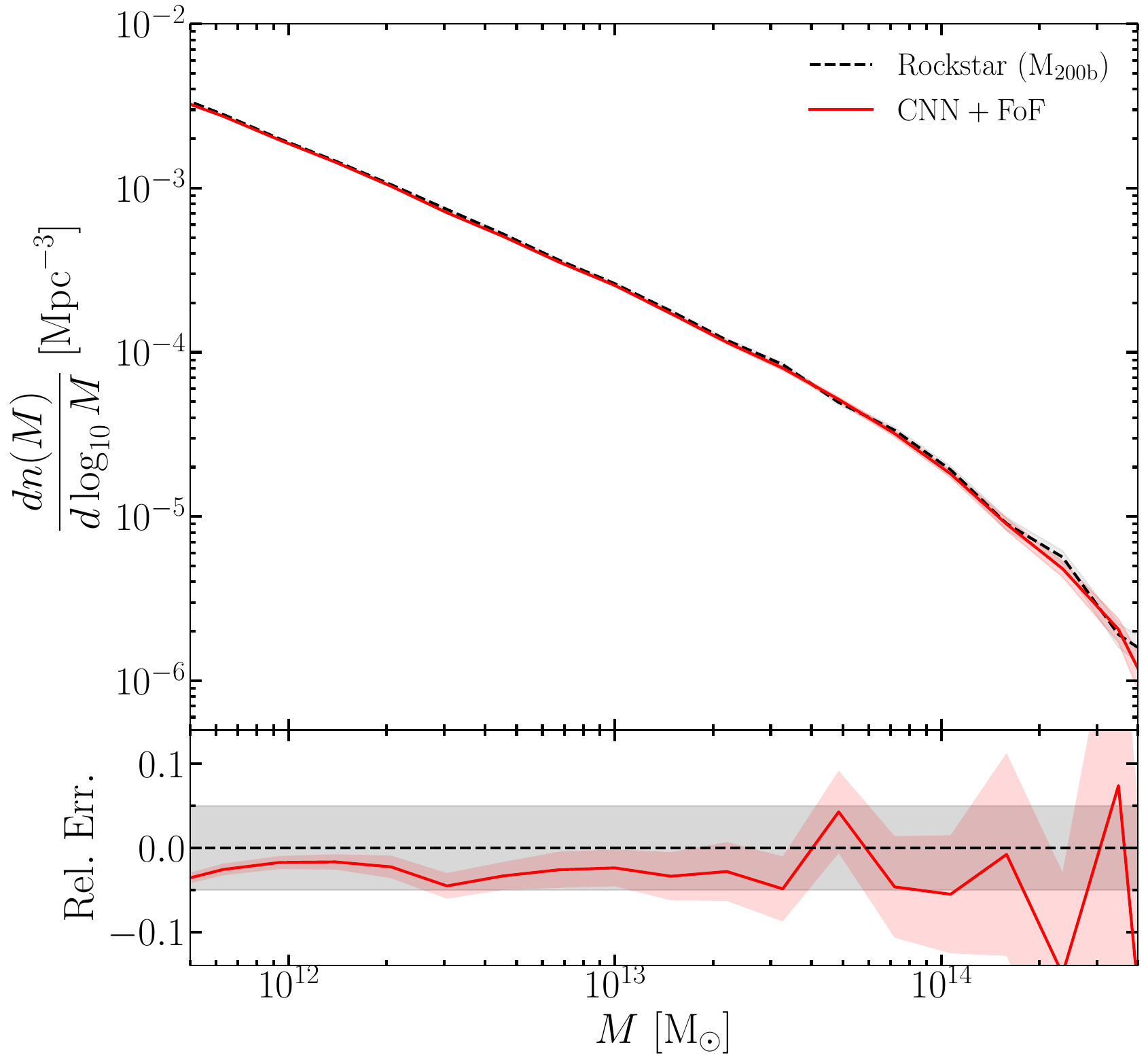}
    \caption[Halo Mass Function Comparison with Relative Error]{
    Comparison of the recovered halo mass function from the \cnnfof{} pipeline (solid red) with the \texttt{ROCKSTAR} reference catalogue (dashed black).
    The bottom panel shows the relative fractional deviation between the two estimates.
    The \cnnfof{} reconstruction agrees within $5\%$ across the majority of the mass range.}
    \label{fig:final_hmf_comparison}
\end{figure}

We now turn our attention to the halo mass function (HMF), a fundamental cosmological statistic.
While the previous comparison focused strictly on cross-matched pairs, the HMF requires the complete halo population; therefore, we compute it using the full samples, including the unmatched haloes discussed earlier.
Figure~\ref{fig:final_hmf_comparison} presents the comoving number density of haloes per logarithmic mass interval, $n(M) = \mathrm{d}N / \mathrm{d}\log M \, \mathrm{d}V$, derived from both the \cnnfof{} pipeline and the reference \texttt{ROCKSTAR} catalogues.
We observe good consistency between the two, with deviations within $5\%$ across the intermediate and high-mass ranges.

\begin{figure}
    \centering
    \includegraphics[width=\columnwidth]{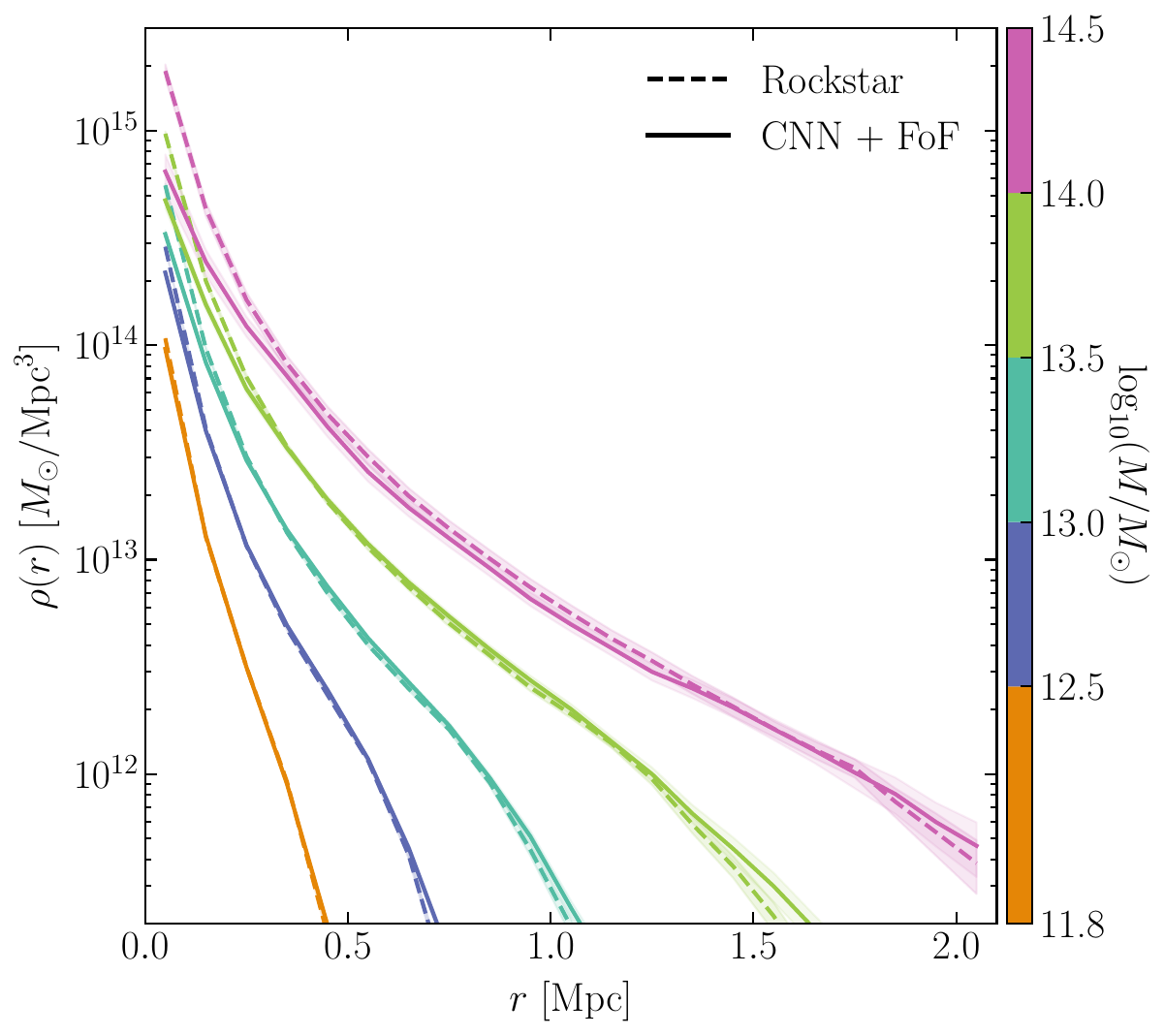}
    \caption[Density Profile]{Mean spherically averaged density profiles of haloes identified by \texttt{ROCKSTAR} (dashed lines) and by the \cnnfof{} pipeline (solid lines). Shaded regions indicate the standard deviation on the mean within each mass bin. The close agreement demonstrates that the \cnnfof{} pipeline accurately reproduces the internal mass distribution of haloes. 
    }
    \label{fig:density_profile}
\end{figure}

We conclude this section by analysing the recovered internal structure of the haloes.
Using the full catalogues, we computed the spherically averaged mass density profiles, $\rho(r)$, for both the \cnnfof{} and the reference \texttt{ROCKSTAR} catalogues.
We divided the samples into five mass bins spanning the entire mass range considered in our work.
The resulting comparisons are presented in Figure~\ref{fig:density_profile}.
The \cnnfof{} profiles closely track the \texttt{ROCKSTAR} results across all mass ranges, demonstrating that our pipeline accurately recovers the detailed internal structure of the haloes.

As a final comment on performance across other resolutions, our baseline analysis focused on the highest-resolution configuration, $L100$-$N128^3$, as this provides the most relevant test for modern cosmological applications by evaluating the network within highly dense, non-linear environments.
We also find the overall performance for the $L200$-$N128^3$ configuration to be statistically consistent with this baseline.
However, while the isolated CNN performs robustly across all tested configurations, the subsequent FoF clustering step systematically degrades the final catalogue quality in simulations with fewer particles, especially in the $L200$-$N32^3$ case.
We attribute this decline to increasing particle sparsity.
Nevertheless, this specific configuration falls far below cosmological standards and served exclusively as a computationally inexpensive environment for rapid calibration.

\subsection{Computational performance}
\label{subsec:computational_performance}

To evaluate the computational speed-up achieved by our method, we compared the total execution time of the \cnnfof{} pipeline against \texttt{ROCKSTAR} across multiple simulation resolutions. It is important to note that this comparison focuses exclusively on the halo-identification stage and does not include the computational cost of generating the fully evolved, non-linear particle field.

\begin{figure}
    \centering
    \includegraphics[width=\linewidth]{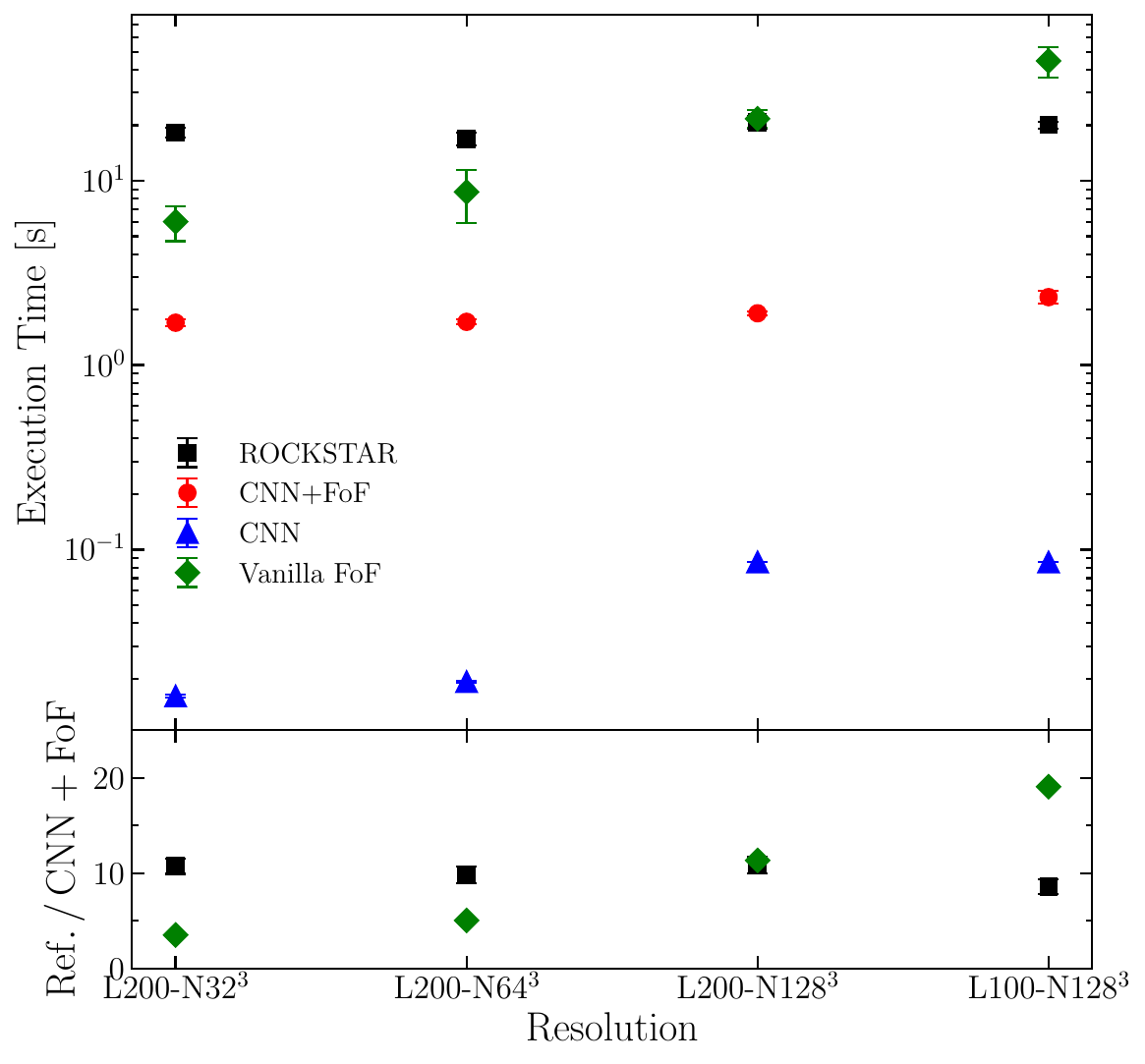}
    \caption[Runtime Comparison of Halo Finders]{
    \textit{Top:} Wall-clock execution time of \texttt{ROCKSTAR} (black squares), the \cnnfof{} pipeline (red circles), the isolated CNN classification stage (blue triangles), and the standard vanilla FoF algorithm applied to the full simulation boxes (green diamonds), across the different resolution configurations. \textit{Bottom:} Ratio of the \texttt{ROCKSTAR} and vanilla FoF execution times to the \cnnfof{} pipeline. The \cnnfof{} pipeline achieves a speed-up of approximately one order of magnitude relative to the reference catalogues.
    }
    \label{fig:runtime_comparison}
\end{figure}

Figure~\ref{fig:runtime_comparison} shows the mean runtime for each method.
The \cnnfof{} pipeline achieves a consistent speed-up of approximately one order of magnitude relative to \texttt{ROCKSTAR}, with runtime ratios ranging from 8 to 12 across all tested resolutions. This improvement arises from the CNN’s ability to infer halo membership in a single, GPU-native forward pass, which is far faster than the iterative six-dimensional phase-space search dominating the cost in \texttt{ROCKSTAR}. The subsequent highly-optimised and parallelised, CPU-based FoF clustering step operates efficiently on the drastically reduced subset of particles classified as halo members, further enhancing overall performance.

To isolate the computational cost of the neural network from the subsequent clustering step, we additionally report the execution time of the CNN classification stage alone.
The CNN operates approximately an order of magnitude faster than the complete \cnnfof{} pipeline across all configurations, remaining stable at $\sim 10^{-1}$~s for the two highest resolutions.
This reveals a critical decomposition of the runtime: particle classification, the most physically complex task, contributes only a minor fraction of the total cost.
Instead, the computational bottleneck remains the FoF clustering stage, despite its highly optimised and parallelised implementation.

For comparison, we also evaluated the execution time of a standard vanilla FoF algorithm applied directly to the full simulation boxes.
While its overall runtimes are comparable to \texttt{ROCKSTAR}, its computational cost increases with resolution, scaling from a factor of $3$ at the lowest resolution to $20$ at the highest.
This degradation reflects the inherent $\mathcal{O}(N^2)$ scaling of the brute-force pairwise distance search as the particle number increases.
Conversely, our parallelised FoF implementation on the pre-filtered particles helps to mitigate this scaling and maintains a more stable performance.

We finish this section with a discussion about our pipeline's strengths and limitations.
First, our pipeline recovers key physical statistics, such as the mass function or the density profile, with a level of precision that is fully competitive with the current state-of-the-art.
To benchmark this performance on similar mass ranges, foundational deep learning models operating on linear initial conditions \citep[e.g.][]{mlhalo_berger_stein,mlhalo_bernardini}, as well as recent differentiable frameworks predicting mass bins on continuous grids \citep[e.g.][]{Buisman2025a}, and state-of-the-art generative emulators \citep[e.g.][]{Pandey_charm2025}, report mass function discrepancies between $~5\%$ and $~15\%$. Achieving a $~5\%$ agreement as in our case, while providing an order-of-magnitude speed-up, demonstrates that our methodology is a robust and viable tool.

The primary limitations of our method arise from the FoF step.
As the classification metrics demonstrate (see Table~\ref{tab:metrics_table}), the isolated CNN performs exceptionally well. However, FoF inherently struggles with ambiguous boundaries and substructures.
While the final \cnnfof{} catalogues achieve high overall purity and completeness (see Table~\ref{tab:matching_table}), the recovery performance fluctuates significantly as a function of mass (as shown in Figure~\ref{fig:halo_matchings}).
This reinforces the fact that the main avenue for future improvement is to find a way to replace or absorb the FoF step entirely into the deep learning architecture.

Another limitation of the method is its dependence on a fixed resolution and redshift.
While the pipeline is in principle scalable to larger volumes, the spatial resolution (particle mass and grid spacing) must remain strictly consistent between the training and inference sets.

Finally, we investigated whether retraining the network with a different minimum number of particles per halo could improve the performance in the intermediate mass regime.
We found that the CNN's classification capabilities are highly stable.
This confirms that the presence of noisy small haloes in the training set does not degrade the network's ability to learn the features of intermediate or high-mass haloes.

\section{Conclusions and future work}
\label{sec:conclusion}

In this study, we have presented a machine-learning-based framework for the identification of dark matter haloes in cosmological simulations. Our methodology combines the strengths of 3D Convolutional Neural Networks with a highly optimised and parallelised FoF algorithm to create a hybrid pipeline (\cnnfof{}) capable of rapidly reconstructing halo catalogues from raw simulation data.

The classification network demonstrated stable performance across multiple simulation resolutions. For the highest-resolution case, $L100-N128^3$, the network consistently achieved over $98\%$ across all primary performance metrics when distinguishing halo and non-halo particles. This particle-level classification served as a robust foundation for the subsequent halo identification. When grouped into individual objects, the resulting catalogue successfully matched $90.54\%$ of the haloes identified by the \texttt{ROCKSTAR} phase-space finder. Spurious detections accounted for a negligible fraction ($0.29\%$) of the total catalogue, indicating minimal contamination.
It is important to note that the discrepancy between the particle-level accuracy ($>98\%$) and the object-level completeness ($\sim 90\%$) does not imply a loss of particles during the FoF stage. The final halo catalogue retains the $98\%$ of correctly classified halo particles. The gap simply reflects that the FoF algorithm groups these particles into distinct objects differently than \texttt{ROCKSTAR}, resulting in an overall $\sim 90\%$ object-to-object match rate.

As a function of halo mass, the purity generally exceeds $99\%$, although it exhibits a mass-dependent decline attributed to FoF fragmentation at low masses and halo merging at the high-mass end.
Completeness stabilises at approximately $94\%$ above $5 \times 10^{11} \, M_{\odot}$, with a drop at the low-mass end arising as we approach the resolution limit.
These results justify the adoption of $5 \times 10^{11} \, M_{\odot}$ as the lower mass limit for the analysis of halo properties.

The spatial and kinematic properties of the recovered haloes exhibited high fidelity relative to the reference catalogue. The normalised positional offsets, $\Delta X_i / r_{200\mathrm{b}}$, were narrowly distributed around zero, and the component-wise velocity ratios were centred near unity with a small scatter. 

The comparison of halo masses reveals that the \cnnfof{} pipeline achieves good agreement with the \texttt{ROCKSTAR} ground truth.
The recovered masses exhibited a tight linear correlation with the reference values, characterised by a mass-dependent scatter that significantly diminishes in the high-mass regime.
This accuracy translates directly into a precise recovery of the halo mass function.
When applying our method to the full catalogues, we reproduced the reference mass function with a consistency of better than $5\%$ across intermediate and high masses.

Beyond global properties, our analysis confirms that the method accurately resolves the internal structure of dark matter haloes.
We examined the spherically averaged density profiles, $\rho(r)$, and found that the \cnnfof{} predictions closely track the \texttt{ROCKSTAR} results across all mass ranges.
This agreement demonstrates that our pipeline not only captures the total mass but also faithfully reconstructs the detailed matter distribution within the haloes.

Several avenues for further development remain. One immediate priority is the integration of the clustering stage directly into the neural network architecture. Replacing the current reliance on a CPU-based FoF step with a fully differentiable, end-to-end halo finder would further streamline the workflow and potentially increase throughput. Additionally, assessing the model's performance across larger simulation volumes and varying cosmological parameters is essential to ensure its generalisability. In this context, techniques such as evolution mapping \citep{Sanchez2022, Esposito2024, Sanchez2026, FiorilliHMF2026} could be leveraged to efficiently transfer learned features across different redshifts and cosmologies, significantly reducing the computational cost of training. 

Ultimately, the primary value of this framework lies in its potential integration into next-generation cosmological analysis tools. By coupling this fast halo-finding pipeline with field-level emulators \citep[e.g.][]{flemu_he, flemu_jamieson} and SBI frameworks, it becomes possible to construct unified and highly efficient forward modelling pipelines. In such frameworks, the rapid identification of haloes from emulated density fields enables the mass-production of mock catalogues necessary for likelihood-free inference, providing a scalable and efficient solution for the high-throughput requirements of modern cosmological surveys.

In summary, our study establishes a robust proof of concept that machine-learning-based halo identification can achieve comparable accuracy to traditional phase-space algorithms while delivering significant computational gains. By bridging the gap between deep learning and classical clustering methods, this hybrid approach provides an efficient alternative for halo identification, well-suited for the analysis of large-scale structure surveys.


\section*{Acknowledgements}

We would like to thank Jiamin Hou and Sofia Contarini for their help and useful discussions. 
Special thanks to Matteo Esposito for his help sharing the tools to process our N-body simulations. We are grateful to the anonymous referee for constructive comments that led to significantly improved final purity, completeness, and mass function metrics of our catalogues, ultimately strengthening the overall results of this work. This analysis was carried out on the HPC system Raven of the Max Planck Computing and Data
Facility (MPCDF, \url{https://www.mpcdf.mpg.de}) in Garching, Germany. This work made extensive use of open-source scientific software, including \texttt{PyTorch}, \texttt{NumPy}, \texttt{SciPy}, \texttt{Matplotlib}, and \texttt{pandas}. 
This research was supported
by the Excellence Cluster ORIGINS, which is funded by the Deutsche
Forschungsgemeinschaft (DFG, German Research Foundation) under Germany's
Excellence Strategy - EXC-2094 - 390783311.
CMC would like to specially thank Daniela Taborda and Martín Correa Taborda for their unconditional support.

\section*{Data Availability}

The data underlying this article will be shared on reasonable request to the corresponding author.


\bibliographystyle{mnras}
\bibliography{bibliography}


\appendix

\section{Confusion matrix and derived metrics}
\label{app:metrics}

To quantitatively evaluate the classification performance of our network, we employ standard metrics derived from the confusion matrix. Let $\mathrm{TP}$, $\mathrm{FP}$, $\mathrm{TN}$, and $\mathrm{FN}$ denote the numbers of true positives, false positives, true negatives, and false negatives, respectively. These represent the counts of correctly and incorrectly classified particles relative to the ground-truth halo catalogue. The total number of positive and negative samples are $P=\mathrm{TP}+\mathrm{FN}$ and $N=\mathrm{TN}+\mathrm{FP}$, respectively, and $T=P+N$ denotes the total number of classified samples.
With this nomenclature, the different metrics are defined as follows:

\begin{align}
\mathrm{Accuracy}      &= \frac{\mathrm{TP}+\mathrm{TN}}{T},\\[3pt]
\mathrm{Precision}     &= \frac{\mathrm{TP}}{\mathrm{TP}+\mathrm{FP}},\\[3pt]
\mathrm{Recall}~(\mathrm{TPR}) &= \frac{\mathrm{TP}}{\mathrm{TP}+\mathrm{FN}},\\[3pt]
\mathrm{Specificity}~(\mathrm{TNR}) &= \frac{\mathrm{TN}}{\mathrm{TN}+\mathrm{FP}},\\[3pt]
\mathrm{F1}            &= \frac{2\,\mathrm{Precision}\times \mathrm{Recall}}{\mathrm{Precision}+\mathrm{Recall}}
= \frac{2\,\mathrm{TP}}{2\,\mathrm{TP}+\mathrm{FP}+\mathrm{FN}},\\[6pt]
\mathrm{MCC}           &= \frac{\mathrm{TP}\cdot \mathrm{TN}-\mathrm{FP}\cdot \mathrm{FN}}
{\sqrt{(\mathrm{TP}+\mathrm{FP})(\mathrm{TP}+\mathrm{FN})(\mathrm{TN}+\mathrm{FP})(\mathrm{TN}+\mathrm{FN})}}.
\end{align}

Accuracy measures the overall fraction of correctly classified particles, but it can be misleading when the data are imbalanced, as is often the case between halo and non-halo particles. Precision quantifies how many of the particles predicted as halo members actually belong to haloes in the reference catalogue, thus penalising false detections. Recall (or true positive rate) measures the completeness, i.e., the fraction of genuine halo particles recovered by the model. Specificity (true negative rate) complements recall by quantifying the fraction of background particles correctly recognised as non-halo. The F1 score provides a harmonic mean between precision and recall, balancing the trade-off between purity and completeness. Finally, the Matthews correlation coefficient (MCC) condenses the four confusion-matrix terms into a single balanced measure, remaining robust even for highly imbalanced datasets.

In addition to these threshold-dependent quantities, we compute threshold-independent metrics by sweeping the decision threshold $\tau\in[0,1]$ across the sigmoid outputs. The area under the ROC curve (AUC) represents the probability that a randomly chosen positive sample is ranked higher than a randomly chosen negative one, serving as an overall measure of separability. The average precision (AP) is the integral of the precision–recall curve, summarising the trade-off between purity and completeness across all thresholds.

In our application, high precision corresponds to few spurious detections (i.e., minimising FP particles), while high recall indicates that the majority of true halo particles are recovered. Metrics such as F1 and MCC, therefore, provide a compact way to assess the balance between purity and completeness of the resulting classification.
Scalar metrics in Table~\ref{tab:metrics_table} are reported at the best classification threshold $\tau$, while AUC and AP summarise the global classification performance across all thresholds.

\section{Performance with the $M_{\mathrm{vir}}$ mass definition}
\label{app:mvir}

Previously, we presented the results corresponding to the halo definition based on $M_{200\mathrm{b}}$ for the L100--N128$^{3}$ high-resolution simulations. Here, we report the analogous results obtained using
the virial mass, $M_{\mathrm{vir}}$, as a reference.

The classification results for $M_{\mathrm{vir}}$ are found to be highly consistent with those obtained for $M_{200\mathrm{b}}$, indicating that the trained model generalises well across different mass definitions. The confusion matrix for this case is presented in Table~\ref{tab:confmat_mvir}, while Table~\ref{tab:metrics_mvir} lists the corresponding performance metrics, including accuracy, precision, recall, and specificity. The values demonstrate that the classification network maintains high performance, confirming its robustness with respect to variations in halo boundary criteria.

\begin{table}
    \centering
    \caption{Confusion matrix for the classification results using the virial mass definition, $M_{\mathrm{vir}}$.}
    \label{tab:confmat_mvir}
    \renewcommand{\arraystretch}{1.3}
    \setlength{\tabcolsep}{12pt}
    \begin{tabular}{l cc}
        \hline
         & \multicolumn{2}{c}{\textbf{Predicted Class}} \\
        \cline{2-3}
        \textbf{True Class} & \textbf{Halo} & \textbf{Non-Halo} \\
        \hline
        \textbf{Halo}      & $23.40\%$ (TP) & $0.87\%$ (FN) \\
        \textbf{Non-Halo}  & $0.89\%$ (FP)  & $75.20\%$ (TN) \\
        \hline
    \end{tabular}
\end{table}

\begin{table}
    \centering
    \caption{Performance metrics for the classification results using $M_{\mathrm{vir}}$.}
    \label{tab:metrics_mvir}
    \renewcommand{\arraystretch}{1.3} 
    \setlength{\tabcolsep}{14pt}      
    \begin{tabular}{lc}
        \hline
        \textbf{Metric} & \textbf{Value} \\
        \hline
        Accuracy    & $0.98$ \\
        Precision   & $0.96$ \\
        Recall      & $0.96$ \\
        Specificity & $0.99$ \\
        F1 Score    & $0.96$ \\
        MCC         & $0.95$ \\
        \hline
    \end{tabular}
\end{table}

The halo matching statistics between the reconstructed catalogue and the \texttt{ROCKSTAR} reference catalogue are summarized in Table~\ref{tab:matching_mvir}. The high matched fraction and low number of unmatched detections confirm the physical consistency of the reconstructed haloes and the reliability of the model when applied to virial mass definitions.

\begin{table}
    \centering
    \caption{Global purity and completeness statistics for the \cnnfof{} halo catalogues using $M_{\mathrm{vir}}$.}
    \label{tab:matching_mvir}
    \renewcommand{\arraystretch}{1.3} 
    \setlength{\tabcolsep}{14pt}      
    \begin{tabular}{lc}
        \hline
        \textbf{Metric} & \textbf{Value} \\
        \hline
        Purity       & $99.18\%$ \\
        Completeness & $86.43\%$ \\
        \hline
    \end{tabular}
\end{table}

Overall, the consistent behaviour for the two mass definitions, $M_{200\mathrm{b}}$ and $M_{\mathrm{vir}}$, demonstrates that the model effectively captures the fundamental physical correlations characterising gravitationally bound structures. Minor variations arise naturally from the different boundary definitions, yet the core predictive capability of the CNN remains robust across all conventions.

\section{Performance with the $M_{\mathrm{FoF}}$ mass definition}
\label{app:mfof}

We additionally report results obtained using the FoF mass definition, $M_{\mathrm{FoF}}$, derived from the standard vanilla FoF algorithm (not our parallelised version).
The confusion matrix is presented in Table~\ref{tab:confmat_mfof}, and the corresponding performance metrics are listed in Table~\ref{tab:metrics_mfof}.
While the overall accuracy and specificity remain relatively high at $0.92$ and $0.95$ respectively, the precision and recall drop to $0.88$ and $0.86$.
Furthermore, both the false-positive ($3.53\%$) and false-negative ($4.43\%$) rates are elevated by roughly an order of magnitude relative to the baseline \texttt{ROCKSTAR} analysis.

In contrast to the spherical-overdensity definitions, this case yields a systematically poorer classification performance.
We attribute this to the highly non-trivial, arbitrary geometries of FoF haloes, which are inherently more difficult for the network to reliably recover.
This test illustrates that ROCKSTAR catalogues constitute a more robust dataset for machine learning recovery than standard FoF catalogues.

It is worth highlighting that, for our baseline \texttt{ROCKSTAR} analysis, we additionally quantified the catalogue-level performance (purity and completeness) by matching the predicted clumps to the reference haloes.
However, applying this procedure to the FoF reference catalogue is not straightforward.
Unlike spherical-overdensity definitions, FoF haloes lack an equivalent notion of size or boundary (such as $r_{200\mathrm{b}}$ or $r_{\mathrm{vir}}$).
Consequently, we omit a direct calculation of global purity and completeness for this specific mass definition.

\begin{table}
    \centering
    \caption{Confusion matrix for the classification results using the FoF 
    mass definition, $M_{\mathrm{FoF}}$.}
    \label{tab:confmat_mfof}
    \renewcommand{\arraystretch}{1.3}
    \setlength{\tabcolsep}{12pt}
    \begin{tabular}{l cc}
        \hline
         & \multicolumn{2}{c}{\textbf{Predicted Class}} \\
        \cline{2-3}
        \textbf{True Class} & \textbf{Halo} & \textbf{Non-Halo} \\
        \hline
        \textbf{Halo}      & $26.15\%$ (TP) & $4.43\%$ (FN) \\
        \textbf{Non-Halo}  & $3.53\%$ (FP)  & $65.90\%$ (TN) \\
        \hline
    \end{tabular}
\end{table}

\begin{table}
    \centering
    \caption{Performance metrics for the classification results using 
    $M_{\mathrm{FoF}}$.}
    \label{tab:metrics_mfof}
    \renewcommand{\arraystretch}{1.3} 
    \setlength{\tabcolsep}{14pt}      
    \begin{tabular}{lc}
        \hline
        \textbf{Metric} & \textbf{Value} \\
        \hline
        Accuracy    & $0.92$ \\
        Precision   & $0.88$ \\
        Recall      & $0.86$ \\
        Specificity & $0.95$ \\
        F1 Score    & $0.87$ \\
        MCC         & $0.81$ \\
        \hline
    \end{tabular}
\end{table}

As a complementary test, we assess how the standard vanilla FoF algorithm performs as a standalone halo finder when applied to the full particle field, in the absence of the CNN pre-selection.
We used a linking-lenght parameter $b = 0.171$ and compared the resulting mass function against the \texttt{ROCKSTAR} reference catalogue. 
This value is the one expected on theoretical grounds for reproducing the spherical-overdensity $M_{200\mathrm{b}}$ mass definition (see Appendix~\ref{subsec:linking_length} for more details).

\begin{figure}
    \centering
    \includegraphics[width=\columnwidth]{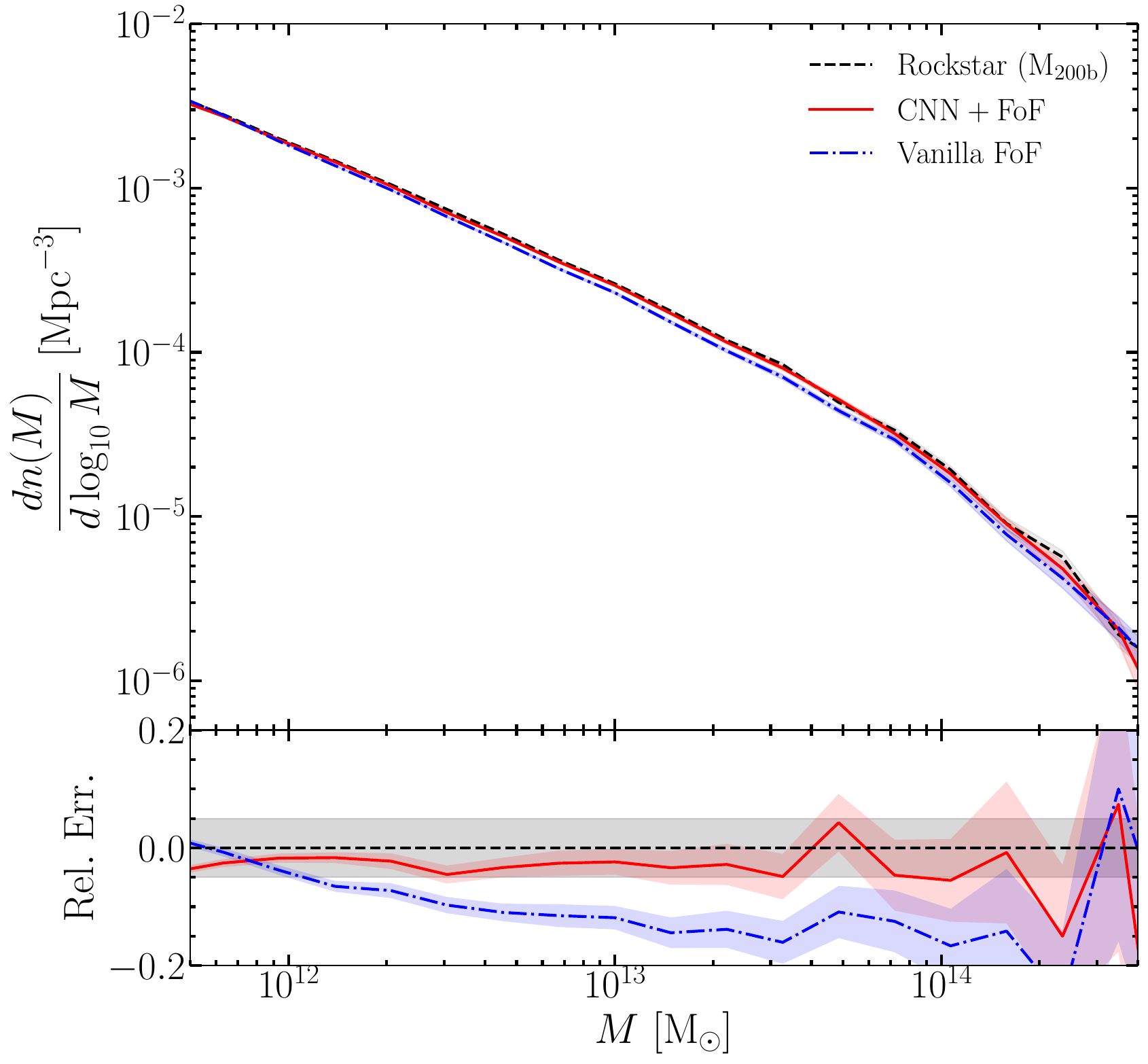}
    \caption{Extension of Figure~\ref{fig:final_hmf_comparison} to include the performance of a vanilla FoF application as a standalone halo finder (dotted-dashed blue). While the \cnnfof{} reconstruction agrees within $5\%$ across the majority of the mass range, vanilla FoF deviates at all masses.
    }
    \label{fig:hmf_vanilla}
\end{figure}

Figure~\ref{fig:hmf_vanilla} shows this comparison with the \cnnfof{} result included alongside.
It is clear that vanilla FoF and \cnnfof{} behave differently when compared to \texttt{ROCKSTAR}.
Our method recovers the reference to within $5\%$ across the resolved mass range, whereas vanilla FoF exhibits a systematic deficit roughly $15-20\%$ across the same mass range, with a deviation that grows towards the high-mass end.
This shows that our method is not only speeding up the process, it is also learning the behaviour of \texttt{ROCKSTAR}.

\section{Calibration of the FoF linking length}
\label{subsec:linking_length}

The FoF linking length is the only free parameter of the clustering stage of our pipeline, and its value governs the balance between the two failure modes of percolation-based grouping. When it is too short, individual objects fragment into spurious sub-clumps. When it is too long, distinct neighbouring structures are bridged into single groups. We can write it as
\begin{equation}
    \lambda = b\,\bar{n}^{-1/3},
    \label{eq:linking_length}
\end{equation}
where $\bar{n}$ is the mean number density of the original full particle field, rather than that of the CNN-selected subset, so that $b$ retains its standard meaning as a fraction of the mean interparticle separation of the simulation.

A natural starting point follows from demanding that percolation terminates at the same overdensity that defines the reference catalogue. Consider a halo satisfying $\bar{\rho}_{\rm h} = \Delta\,\rho_{\rm b}$, where $\rho_{\rm h}$ is the mean mass density enclosed within the halo boundary, $\rho_{\rm b}$ is the comoving background matter density of the simulation, and $\Delta$ is the overdensity threshold that specifies a given mass definition. 
Since all particles carry the same mass, this condition translates directly into a number density $n_{\rm h} = \Delta\,\bar{n}$, and therefore, the characteristic interparticle separation can be written as
\begin{equation}
    \ell = n_{\rm h}^{-1/3} = \Delta^{-1/3}\,\bar{n}^{-1/3}.
\end{equation}
Setting the linking length equal to this separation, $\lambda = \ell$, gives
\begin{equation}
    b = \Delta^{-1/3},
    \label{eq:b_theory}
\end{equation}
so that particles link only where the local sampling is at least as dense as that of a region at the target overdensity.
In particular, for the $M_{200\mathrm{b}}$ definition adopted for our reference catalogue, $\Delta = 200$, and equation~(\ref{eq:b_theory}) yields $b = 0.171$.
This is the value used for the standalone vanilla-FoF comparison of Appendix~\ref{app:mfof}.

\begin{figure*}
    \centering
    \includegraphics[width=\textwidth]{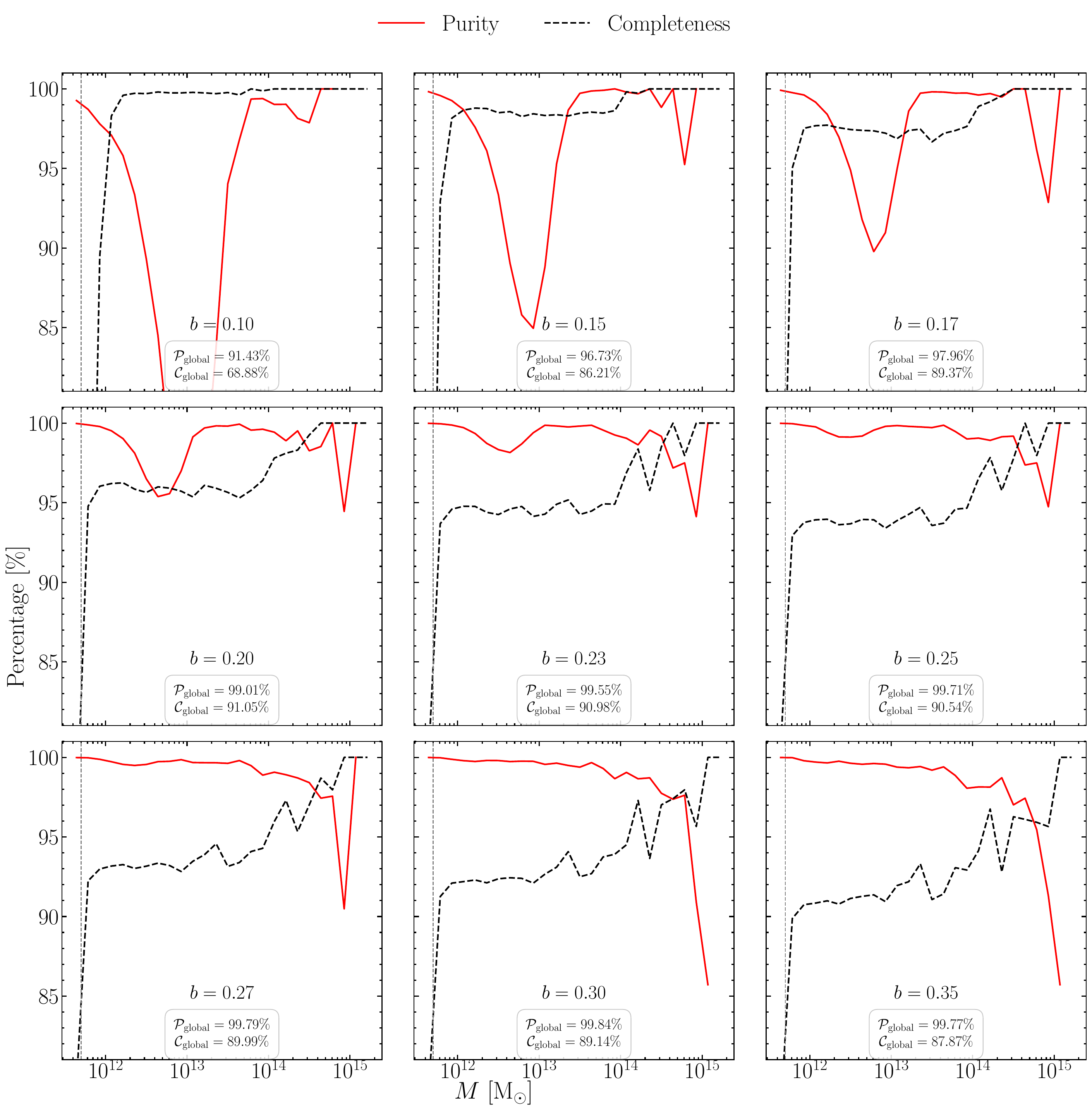}
    \caption{Mass-dependent purity (solid red lines) and completeness (dashed black lines) of the \cnnfof{} catalogues for nine values of the FoF linking-length parameter $b$, evaluated on the $L100$-$N128^3$ test set against the \texttt{ROCKSTAR} $M_{200\mathrm{b}}$ reference. Each panel quotes the global purity, $\mathcal{P}_{\rm global}$, and completeness, $\mathcal{C}_{\rm global}$, computed over the full mass range. As $b$ increases, we observe a trade-off as a function of mass: while purity systematically improves, completeness behaves in the opposite manner. Recovery stabilises into an optimal operational range between $b = 0.20$ and $b = 0.27$.
    }
    \label{fig:pc_bscan}
\end{figure*}

\begin{figure*}
    \centering
    \includegraphics[width=\textwidth]{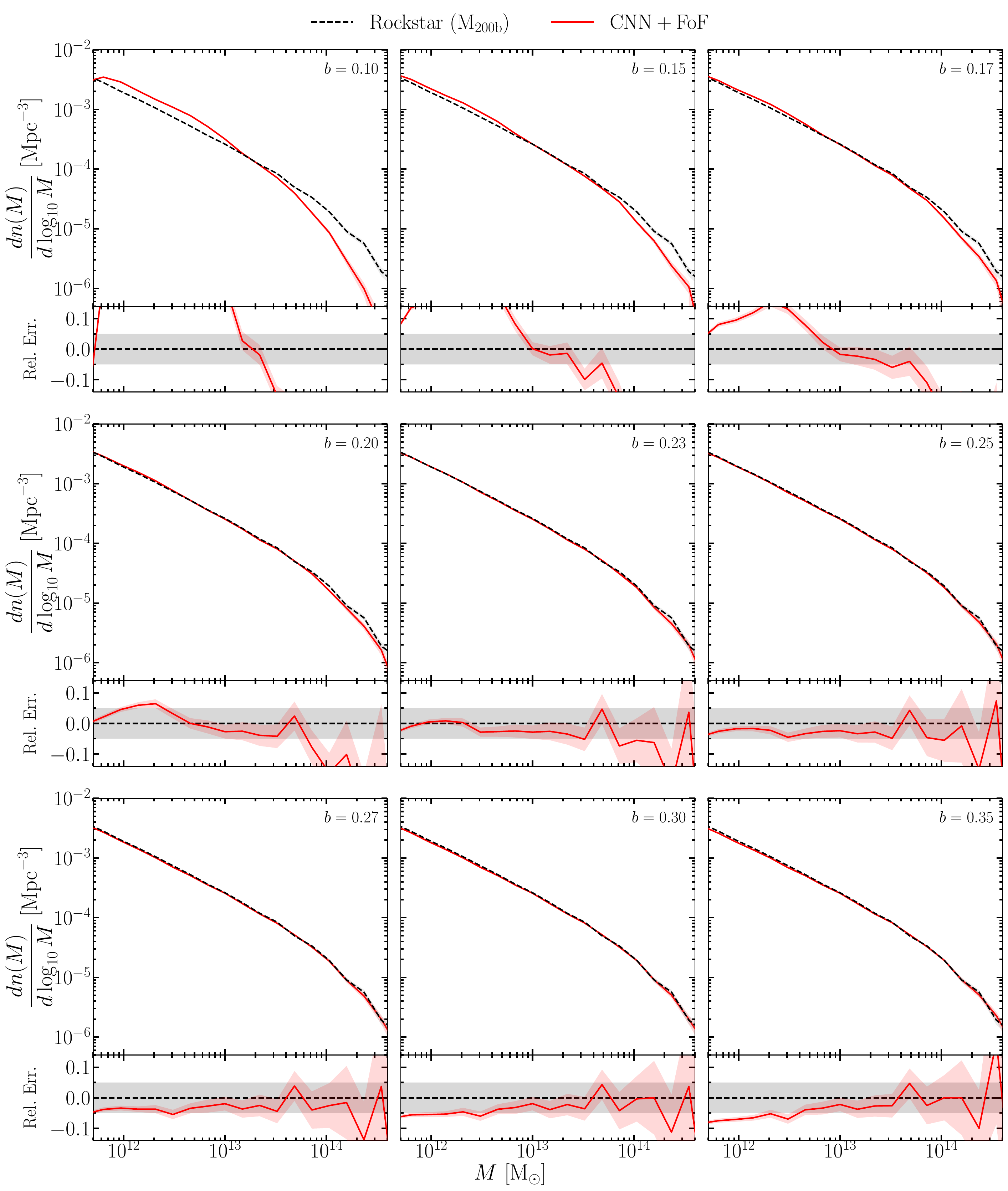}
    \caption{Halo mass functions recovered by the \cnnfof{} pipeline (solid red lines) for the nine values of the FoF linking-length parameter $b$, compared with the \texttt{ROCKSTAR} $M_{200\mathrm{b}}$ reference (dashed black). The lower sub-panels show the relative deviations from the reference. Fragmentation at low values of $b$ tilts the mass function, producing a low-mass excess and a high-mass deficit, whereas largeer values yield a low-mass deficit caused by the merging of small systems into more massive neighbours. The fiducial choice $b = 0.25$ keeps the residuals small across the resolved mass range.
    }
    \label{fig:hmf_bscan}
\end{figure*}

For an empirical calibration of $b$, we processed the CNN-filtered test set by scanning values in the range $b \in \{0.10, 0.15, 0.17, 0.20, 0.23, 0.25, 0.27, 0.30, 0.35\}$, keeping every other element of the pipeline fixed, and then evaluated the resulting catalogues against the \texttt{ROCKSTAR} $M_{200\mathrm{b}}$ reference following the matching procedure of Section~\ref{subsec:recovered_halos}.
This section shows the results for the $L100$-$N128^3$ configuration, as this is the reference used throughout the main body of the text.
In Appendix~\ref{app:resolutions}, we report the corresponding results for other resolutions.
Figure~\ref{fig:pc_bscan} shows the mass-dependent purity and completeness for each value of $b$, with the corresponding global statistics quoted in each panel, whereas Figure~\ref{fig:hmf_bscan} compares the recovered mass functions against the reference.

The parameter scan separates two regimes.
At lower values of $b$, the catalogue is dominated by fragmentation, where massive systems are artificially split into several smaller clumps.
Consequently, the mass function acquires a characteristic tilt, showing an excess of low- and intermediate-mass objects compensated by a deficit at the high-mass end.
As $b$ increases, we observe a clear trade-off between our key metrics as a function of mass: while purity systematically improves with larger linking lengths, completeness behaves in the opposite manner, deteriorating as $b$ grows.
We observe that recovery stabilises into an optimal operational range between $b = 0.20$ and $b = 0.27$.
Throughout the main body of this work, we adopt $b = 0.25$ from within this window, as it provides the best overall balance across the considered mass range, attaining a global purity of $99.71\%$, a completeness of $90.54\%$, and recovering the halo mass function well within a $5\%$ error band.

Finally, we note that the results favour the value $b = 0.25$ over the expected theoretical one given by $b = 0.171$.
A possible explanation can be attributed to the fact that FoF tends to underestimate the mass compared to \texttt{ROCKSTAR}, as can be appreciated in Figure~\ref{fig:hmf_vanilla}.
Hence, it is reasonable to assume a larger linking length for this particular task.

\section{Performance across simulation resolutions}
\label{app:resolutions}

\begin{figure*}
    \centering
    \includegraphics[width=\textwidth]{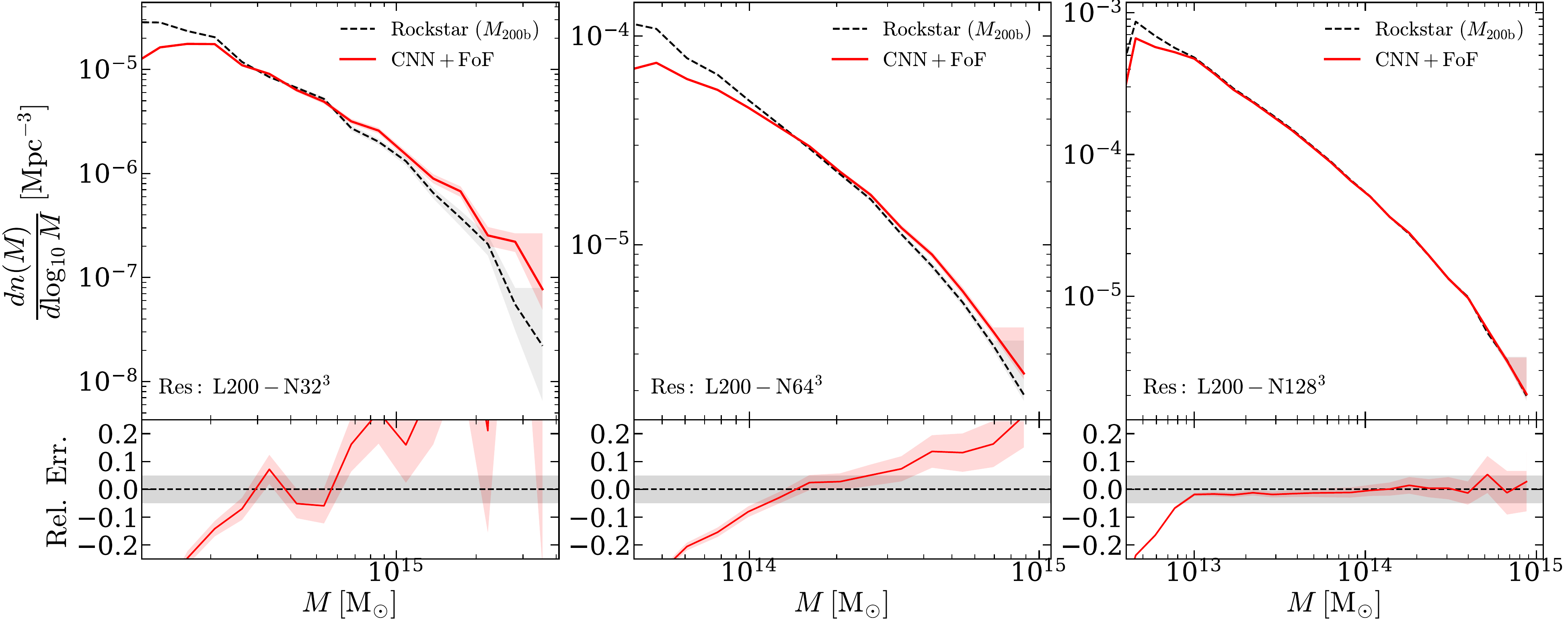}
    \caption{Halo mass functions recovered by the \cnnfof{} pipeline (solid red lines) compared with the \texttt{ROCKSTAR} $M_{200\mathrm{b}}$ reference (dashed black) for the three $L200$ configurations of Table~\ref{tab:mass_table}, namely $L200$-$N32^3$ (left), $L200$-$N64^3$ (centre) and $L200$-$N128^3$ (right). The equivalent result for $L100$-$N128^3$ is shown in Figure~\ref{fig:final_hmf_comparison}.
    The overall performance of the $L200$-$N128^3$ configuration is statistically consistent with the $L100$-$N128^3$ baseline, the primary and highest resolutions considered in this work.
    While the isolated CNN performs robustly across all configurations, the subsequent FoF clustering step systematically degrades the final catalogue quality in simulations with fewer particles.
    }
    \label{fig:hmf_res}
\end{figure*}

The analysis of the main text focuses on the highest-resolution configuration, $L100$-$N128^3$, whose recovered mass function is shown in Figure~\ref{fig:final_hmf_comparison}. In order to test the performance across different resolutions, we present here the corresponding $M_{200\mathrm{b}}$ mass functions for the three remaining configurations of Table~\ref{tab:mass_table}, each compared against its own \texttt{ROCKSTAR} reference. These are shown in Figure~\ref{fig:hmf_res}. The global purity and completeness of all four configurations are reported in Table~\ref{tab:matching_table}, and the particle-level classification metrics in Table~\ref{tab:metrics_table}, so that the full set of resolution-dependent diagnostics can be read together.

For each configuration, we repeated the linking-length scan test of Appendix~\ref{subsec:linking_length}.
The empirically preferred values are $b = 0.23$ for $L200$-$N128^3$, $b = 0.33$ for $L200$-$N64^3$ and $b = 0.49$ for $L200$-$N32^3$.
The larger values required by the coarser boxes reflect a dependency on the fraction of filtered particles provided by the CNN: from approximately $30\%$ to $14\%$ and $7\%$, respectively.

We find the overall performance of the $L200$-$N128^3$ configuration to be statistically consistent with our $L100$-$N128^3$ baseline, so the pipeline is not tied to a single box size. 
While the isolated CNN performs robustly across all tested configurations, the subsequent FoF clustering step systematically degrades the final catalogue quality in simulations with fewer particles.
The $L200$-$N64^3$ configuration exhibits a transitional decline, and the $L200$-$N32^3$ case shows a more significant deviation.
We attribute this decline to increasing particle sparsity.
Nevertheless, this last case falls far below the standards required for cosmological applications and serves exclusively as a computationally inexpensive environment for rapid calibration.

Finally, the tests presented here show that $b$ must be recalibrated for the particular resolution to be used as it acts as an effective parameter.
Applying the pipeline without any reference catalogue would require a predictive relation for $b$, which is out of the scope of this work.

\bsp
\label{lastpage}
\end{document}